\documentclass{nature}

\usepackage{epsfig}
\usepackage{color}
\newcommand{\gtrsim}{\lower.7ex\hbox{$\;\stackrel{\textstyle>}{\sim}\;$}}
\newcommand{\lesssim}{\lower.7ex\hbox{$\;\stackrel{\textstyle<}{\sim}\;$}}

\title{A measurement of the intrahalo light fraction with near-infrared background anisotropies}

\author{Asantha Cooray$^1$, Joseph Smidt$^1$, Francesco De Bernardis$^1$, Yan Gong$^1$, Daniel Stern$^{2}$, Matthew L. N. Ashby$^3$, Peter R. Eisenhardt$^2$, 
Christopher C. Frazer$^1$, Anthony H. Gonzalez$^4$, Christopher S. Kochanek$^5$, Szymon Koz{\l}owski$^{6,5}$ \& Edward L. Wright$^7$ 
}

\begin{document}

\maketitle

\begin{affiliations}
\item Dept. of Physics \& Astronomy, University of California, Irvine, CA 92697, USA
\item Jet Propulsion Laboratory, California Institute of Technology, Pasadena, CA 91109, USA
\item Harvard-Smithsonian Center for Astrophysics, 60 Garden St., Cambridge, MA 02138, USA
\item Department of Astronomy, University of Florida, Gainesville, FL 32611, USA
\item Department of Astronomy, The Ohio State University, Columbus, OH 43210, USA
\item Warsaw University Observatory, Al. Ujazdowskie 4, 00-478 Warszawa, Poland
\item Department of Physics and Astronomy, University of California, Los Angeles, CA 90095, USA
\end{affiliations}


\begin{abstract}
Unresolved near-infrared background anisotropies are expected to have contributions from the earliest galaxies 
during reionization\cite{Santos02,Salvaterra03,Cooray04,Kashlinsky04,Fernandez10} and faint, dwarf galaxies at intermediate redshifts\cite{Chary08,Helgason12}.
Previous measurements\cite{Kashlinsky05,Cooray07,Matsumoto11,Thompson08,Kashlinsky12} were unable to
conclusively pinpoint the dominant origin because they did not sample spatial scales that were sufficiently large to distinguish between these two 
possibilities. Here we report a measurement of the anisotropy power spectrum from sub-arcminute to one degree angular scales
and find the clustering amplitude to be larger than the model predictions involving the two existing explanations.
As the shot-noise level of the power spectrum is consistent with that expected from faint galaxies,
 a new source population on the sky is not necessary to explain the observations. A physical mechanism that increases the
clustering amplitude, however, is needed. Motivated by recent results related to the extended stellar light profile in dark matter 
halos\cite{Purcell07,Purcell08,Conroyetal07}, we consider the possibility that the fluctuations originate from
diffuse intrahalo stars of all galaxies. We find that the measured power spectrum can be explained by an intrahalo light fraction of 0.07 to 0.2\% relative
to the total luminosity  in dark matter halos of 10$^{9}$ to 10$^{12}$ solar masses at redshifts of $\sim$ 1 to 4. 
\end{abstract}

In order to distinguish between the two interpretations of the  near-IR anisotropy power spectrum,
we have analyzed imaging data from the {\it Spitzer} Deep, Wide-Field Survey (SDWFS)\cite{Ashby:2009}.
This survey covers 10.5 square degrees on the sky with the IRAC instrument in its four bands between 3.6 and 8 $\mu$m. 
We focus on the data at 3.6 and 4.5 microns as the confusion from zodiacal light limits extragalactic background studies
at 5 and 8 microns\cite{Kashlinsky05}.  The data were taken in four separate epochs between 2004 and 2008
and were conducted in ways to minimize the systematics associated with anisotropy measurements.
In particular the  four different epochs were observed at different roll angles of the instrument so that
the measurements are robust against detector artifacts,
persistence resulting from saturated bright stars, and variations in the bias level.
The SDWFS mapping strategy was also optimized to 
facilitate self-calibration\cite{Arendt} of the data by maximizing inter-pixel correlations.

To limit the influence of bright stars and galaxies, including extended sources, in our anisotropy measurements  we mask 
all sources that are detected either in the combined SDWFS data or in the ancillary multi-band optical and near-IR data\cite{JannuziDey99}.
The effects of mosaicing  of individual detector frames, pixelization of the maps, and the detected-source mask are
captured by the map-making transfer function (see Supplementary Information). We compute the transfer function and its uncertainty
with a large set of sky simulations. The point-spread function (PSF) and 
its uncertainty were determined by measuring and modeling the PSF of stars at different sub-regions
of the image and computing the variance of the differences between the modeled PSFs.

The power spectrum measurements at 3.6 and 4.5 $\mu$m show a clear excess above the shot-noise level
(Fig.~1).  The shot-noise dominates the anisotropy power spectrum at sub-arcminute angular scales corresponding to $\ell > 10^5$.
Such a shot-noise is expected from the small-scale Poisson behavior of the spatial distribution of sources on the sky.
The clustering amplitude we measure at $\ell \sim 10^4$ is fully consistent with existing
measurements of the anisotropy power spectrum with IRAC in deeper, but smaller area, fields\cite{Kashlinsky05,Cooray07,Kashlinsky12}.  At tens of arcseconds angular scales,
corresponding to $\ell > 5\times10^4$, our shot-noise level is higher than that of a recent measurement by about a factor of 2 because deeper data  
allow more faint sources to be individually detected and masked\cite{Kashlinsky12}.  Nevertheless, we independently confirm
the near-IR background anisotropies at angular scales larger than a few arcminutes at the previously reported amplitude\cite{Kashlinsky12}.

The near-IR anisotropies have been previously interpreted as either due to spatial clustering of primordial galaxies responsible for cosmic reionization\cite{Kashlinsky05} or due to faint, 
dwarf galaxies at low redshifts that fall below the individual source detection threshold of {\it Spitzer} images\cite{Cooray07,Chary08,Helgason12}.  Fig.~1 shows that the measured fluctuations in SDWFS are well above both 
these interpretations.
The power spectrum predictions for $z > 6$ galaxies rely on a combination of analytical calculations\cite{Cooray12} and numerical simulations\cite{Fernandez12} of reionization.
If we force the $z>6$ galaxy model to fit the power spectrum data then the integrated intensity of $z>6$ galaxies is 
about 2 nW m$^{-2}$ sr$^{-1}$ at 3.6 $\mu$m\cite{Cooray12}.
In order to reach such a high intensity these galaxies must be very efficient in converting baryons to stars\cite{Fernandez12}. 
In fact, the required star-formation rate conflicts with the measured metal abundance at $z>4$, the measured X-ray background when compared to X-rays from stellar end products such as black holes,
and the measured luminosity functions of bright Lyman-dropout galaxies\cite{MadauSilk}. Unless a significant revision of our current understanding of $z>6$ galaxy statistics is made
it is unlikely that the measured anisotropy power spectrum is dominated by the primordial galaxies.

The prediction for low redshift, faint galaxy intensity fluctuations involves 
a large compilation of multi-wavelength luminosity functions and galaxy number counts\cite{Helgason12}.
The measured luminosity function slope at the faint-end is used to extrapolate to the fainter galaxies that are undetected in the {\it Spitzer} images.   
An increase in the faint-end slope above the measured values does not increase the clustering amplitude on the angular scales of interest without modifying the 
shot-noise level.
While the clustering amplitude is smaller than the measurements,  the prediction related to faint, dwarf galaxies\cite{Helgason12} shows that they generate 
a shot-noise level consistent with the measured small-scale anisotropy power spectrum (Fig.~1).  
At few tens arcminute angular scales the measurements are such that the clustering amplitude is about a factor of 6 to 10 above the prediction. 
While this difference suggests that a new model to explain the anisotropy power spectrum is clearly needed, the 
consistency with the shot-noise level is such that we do not need to invoke  a new population of point sources on the sky to explain the observations.

While keeping the shot-noise level the same, the measurements can be explained by any physical effect that boosts the two-halo term of clustering.
One possibility is to increase the halo mass scale of the faint, dwarf galaxies so that clustering bias factor is increased. The required modification needed
to explain the fluctuations data, however, is ruled out by the measured number counts and the redshift distribution\cite{Helgason12}.
Since intensity anisotropies are measured, another option is to  introduce a luminosity component
to the dark matter halos that remain unmasked when the hosted bright galactic disks are masked as part of the analysis.
Such a possibility exists in the literature in the form of diffuse halo stars in the
extended stellar profile of galaxies out to distances of 100 kpc\cite{TalVanDokkum11}.
In our anisotropy measurements, we mask the faintest detected galaxies to 3-4$''$ which removes  the light from the bulges and disks of those galaxies.
To remove the diffuse light component we would have to mask to a radius greater than  $10''$ around each galaxy. The
surface density of galaxies down to $m_{\rm AB} < 22$ at 3.6 $\mu$m is such that we expect 2 to 3 galaxies
within a circle of radius 10$''$. Thus masks which successfully remove the diffuse component leave no pixels on the map from which to measure the anisotropy power spectrum.

Existing studies discuss this extended emission in terms of the diffuse intrahalo light (IHL)\cite{Conroyetal07} and explain the origin through
tidally stripped stars during galaxy mergers and collisions. The stripped fraction is expected to be a function of the halo mass 
with more massive halos containing a larger fraction  of the diffuse halo emission\cite{LinMohr04,Purcell07,Purcell08}.
On galaxy cluster scales the diffuse intracluster light\cite{Rudick09,Gonzalez04} is a significant  fraction of the  total luminosity of the cluster. 
We describe the  intensity anisotropy power spectrum from the IHL by modifying the standard galaxy clustering predictions\cite{CooraySheth02} to include a profile
for the diffuse stars in halos (see Section 8 of the Supplementary Information).  
If the clustering excess in near-IR anisotropy power spectrum is due to IHL, then we find
 that measured anisotropies can be described with halos in the mass range of 10$^9$ to 10$^{12}$ M$_{\odot}$.
Averaged over this mass range we find an $f_{\rm IHL}$ of 0.07 to 0.2\% at 68\% confidence level (Fig.~2). The implied fraction is consistent with the
theoretical expectation that the IHL level is small for low mass halos, but differences also exist with current theory predictions\cite{Purcell07,Purcell08}, especially
in terms of the power-law slope of the halo mass dependence. 

If this new interpretation involving IHL is the correct description of measured IR background anisotropies,  we find that the IHL in all dark matter halos that we are probing
contribute $0.75 \pm 0.25$ nW m$^{-2}$ sr$^{-1}$ to the total intensity at 3.6 $\mu$m. This intensity can be compared to the rms fluctuations
of about 0.1 nW m$^{-2}$ sr$^{-1}$ at a few arcminutes angular scale  (see Fig.~3). The IHL fluctuation signal
varies spatially at a level of 10 to 15\% of its integrated intensity and below 1\% of the total background intensity 
of 13.3 $\pm$ 2.8 nW m$^{-2}$ sr$^{-1}$ at 3.6 $\mu$m\cite{Levenson07}. As the spectral energy distribution of IHL is mostly unknown, we make use of
a variety of SEDs from B to K-type stellar spectral templates  and find an order of magnitude variation at wavelengths of $\sim$ 1 $\mu$m (Fig.~3). In all these cases
we predict the existence of optical background light fluctuations. They will have a similar power spectrum shape and will be fully correlated with fluctuations at 3.6 $\mu$m.
Furthermore the near-IR anisotropies we have measured should be correlated with the
sub-mm anisotropies\cite{Amblard11}, especially if there are diffuse and extended dust associated with galaxies. These form future tests that can be
used to improve our understanding of the content and nature of IHL in distant dark matter halos.


\begin{addendum}
 \item 
We acknowledges support from NSF CAREER (to A.C.), NASA ADAP, and an award issued by JPL/Caltech.
We thank R.~Arendt for sharing his IRAC self-calibration code. We thank J. Bock and M. Zemcov for their contributions to the SDWFS project.
This work is based on observations made with the Spitzer Space Telescope. This work also made use of data products provided by the NOAO Deep Wide-Field Survey. 
A.C. thanks the Aspen Center for Physics for hospitality.

\item[Author Contributions] A.C. planned the study, developed the intrahalo light model, supervised the research work of J.S., F.D. C.F. and 
Y.G.,  and wrote the draft version of this paper.
J.S. and C.F. 
performed the power spectrum measurements and F.D. interpreted those measurements with a halo model for the intrahalo light. Y.G. developed a model for the high-redshift
galaxies. All other coauthors of this paper contributed extensively and equally by their varied contributions to the SDWFS project (led by D.S. as the PI),
planning of SDWFS observations, analysis  of SDWFS data, and by commenting on this manuscript as part of an internal review process.

 \item[Correspondence] Correspondence and requests for materials should be addressed to A.C. (acooray@uci.edu)

 \item[Competing interests statement] The authors declare no    competing interests.
 \item[Supplementary information] accompanies this paper.

\end{addendum}

\clearpage
\newpage


\begin{figure}
\centerline{
\includegraphics[width=8cm,clip]{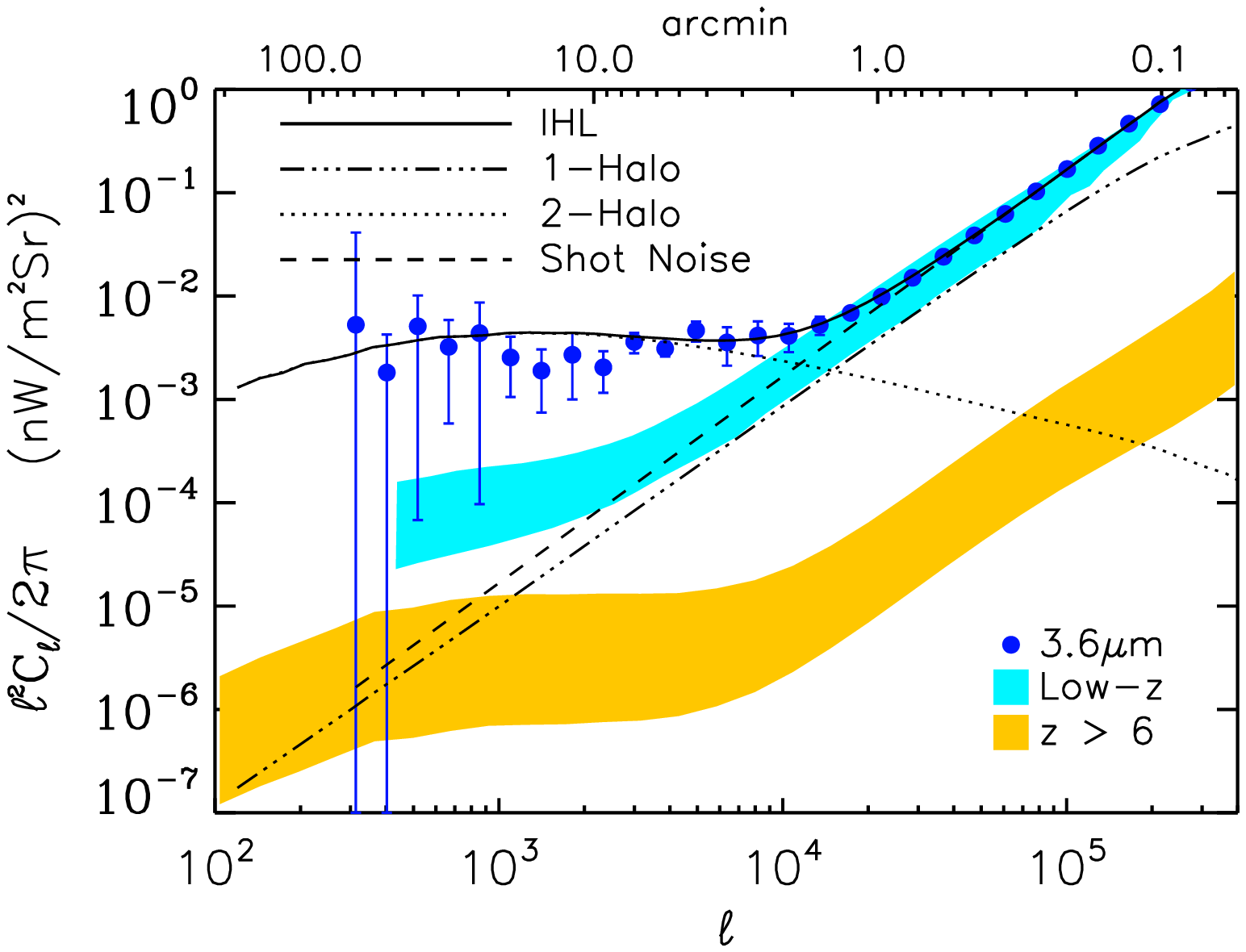}
\includegraphics[width=8cm,clip]{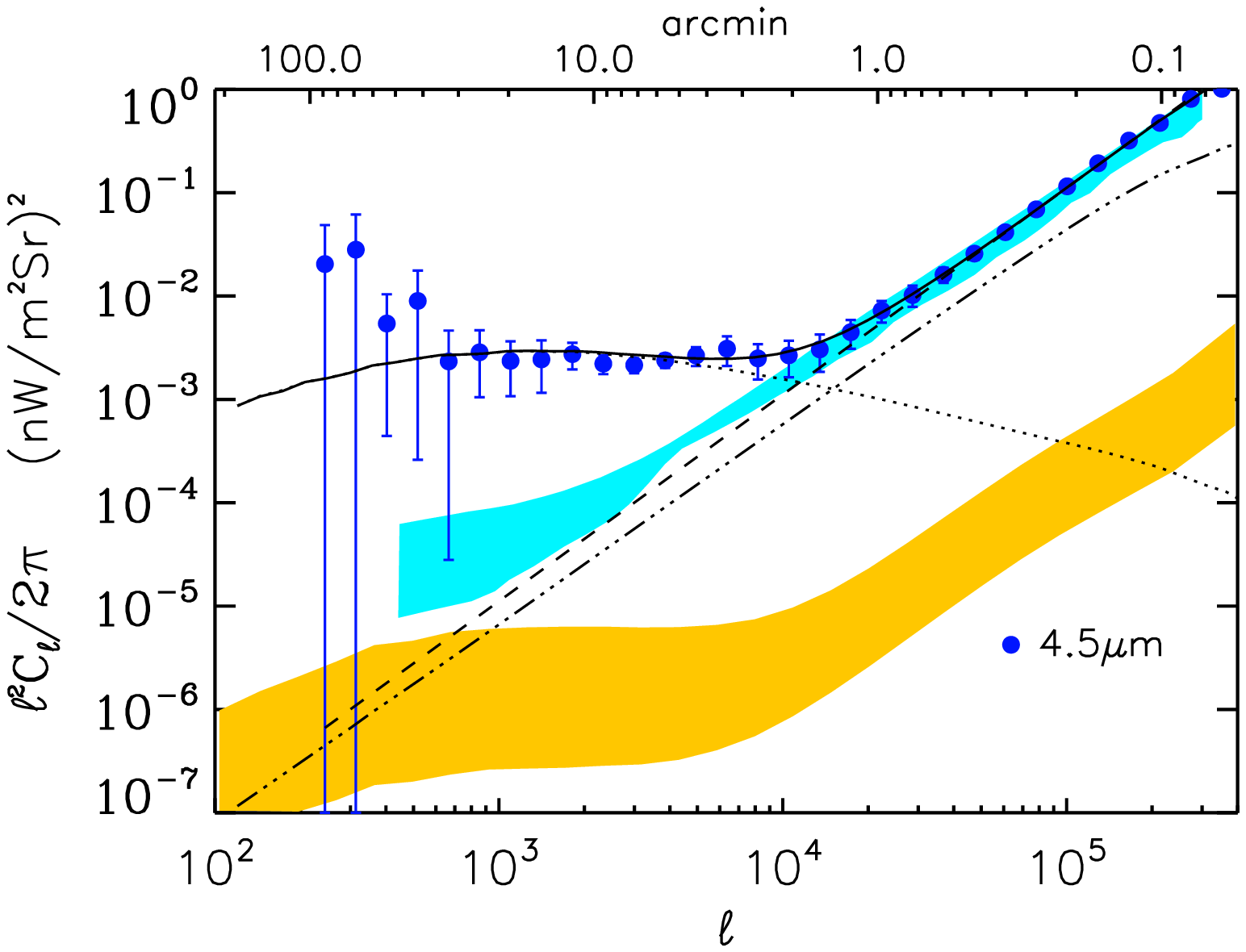}
}
\caption{{\bf The angular power spectrum of the unresolved near-IR background}.
The total power spectrum $P(k)$ of SDWFS at 3.6 $\mu$m (top) and 4.5 $\mu$m (bottom) as a function of the multipole moment. 
SDWFS imaging data were taken on the same field at four separate epochs in January 2004, August 2007, February 2008, and March 2008. 
Each epoch of data, taken over 7 to 10 days,  includes 4300 to 4900 IRAC frames that were combined to make mosaics using the self-calibration algorithm\cite{Arendt}.
The total integration time is 6 minutes per pixel.  These individual frames were first visually inspected and cleaned of artifacts such as asteroidal trails and hot pixels. 
Through cross-correlations between sum and difference maps between epochs, we make independent measurements of the sky signal and noise.
The final power spectrum is the average of the multi-epoch cross-correlation data under the assumption that the instrumental noise is not correlated between epochs. 
In both panels the error bars are 1$\sigma$ uncertainties in  the power spectrum. They are determined by
propagating the errors from the beam measurement into the power spectrum, while the simulations, based out of noise measurements, were used to obtain
instrumental and sky variance.  The quadratic sum of these errors and the map-making transfer function uncertainty constitutes the final error estimate. 
The two shaded regions show the expected contribution from $z > 6$ galaxies\cite{Cooray12} and low-redshift galaxies\cite{Helgason12} based on two model predictions in the literature.
The lines shows a diffuse intrahalo light model where we show the signal in terms of the total (solid), one (dashed-dotted) and two (dotted) halo terms.
The dashed line is the best-fit shot-noise signal that dominates the anisotropies at small angular scales.
}
\label{Cl}
\end{figure}

\clearpage

\begin{figure}
\centerline{\includegraphics[width=11cm,clip]{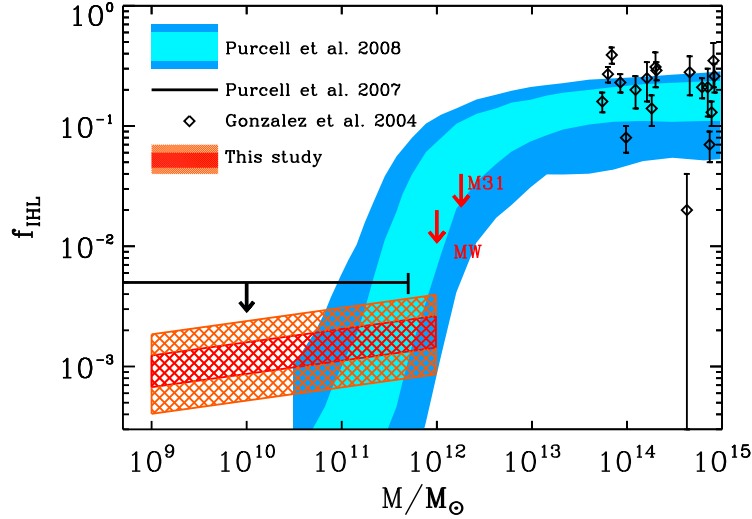}}
\caption{{\bf The intrahalo light fraction from diffuse stars as a function of the halo mass.} 
 The dark and light blue shaded regions show the 95\% and 68\% range of $f_{\rm IHL}$ relative to the total luminosity of the dark matter halos as a function of the halo mass from an analytical prediction\cite{Purcell08}, valid for $f_{\rm IHL}>4\times10^{-4}$ and $M > 5 \times 10^{10}$ M$_{\odot}$ and at $z= 0$.  
We show the case where subhalos on orbits passing within a critical radius of the host halo center contribute their light 
to the central galaxy rather than to the diffuse component. We also show a prediction where $f_{\rm IHL}$ is constant\cite{Purcell07},
due to dwarf galaxies that are completely destroyed, with a value $\sim 0.005$ when $M \lesssim 5\times10^{11}$ M$_{\odot}$ 
(solid line fixed at  $f_{\rm IHL}=5\times10^{-3}$). The downward arrow indicates the possibility that the constant $f_{\rm IHL}$ value 
for low mass halos may be smaller at higher redshifts. 
The red and orange hatched regions at the bottom of the plot are the preferred 68\%  and 95\% confidence level range, respectively, on $f_{\rm IHL}$ from our
analysis of the SDWFS near-IR anisotropy power spectrum.  The mass range is determined by the minimum and maximum halos masses consistent with the
halo model fit that includes the IHL component. Both the mass and $f_{\rm IHL}$ ranges are valid over the broad redshift interval from $z=1$ to 4 
over which the anisotropy signal is generated. We do not find a significant halo mass dependence on the 
IHL fraction with the mass-dependent power-law to be $0.09 \pm 0.01$ between 10$^9$ to 10$^{12}$ M$_{\odot}$,
consistent with the possibility that $f_{\rm IHL}$  is mass independent\cite{Purcell07} when  $M \lesssim 5\times10^{11}$ M$_{\odot}$.
Our model requires the total luminosity-halo mass relation to evolve with redshift as $(1+z)^{1.2 \pm 0.1}$. This 
luminosity evolution with redshift can also be absorbed into the evolution of $f_{\rm IHL}(M)$ evolution with redshift. 
For reference, we also show measurements and 1$\sigma$ errors of the  intracluster light\cite{Gonzalez04}, the
galaxy group and cluster analog for IHL when $M > 5 \times 10^{13}$ M$_{\odot}$.  At halo masses around $10^{12}$ M$_{\odot}$
we show the 95\% confidence level upper limit on $f_{\rm IHL}$ estimated for Milky Way\cite{Carollo10} and Andromeda (M31)\cite{Courteau10}. 
}
\label{fIHL}
\end{figure}

\clearpage

\begin{figure}
\centerline{\includegraphics[width=11cm,clip]{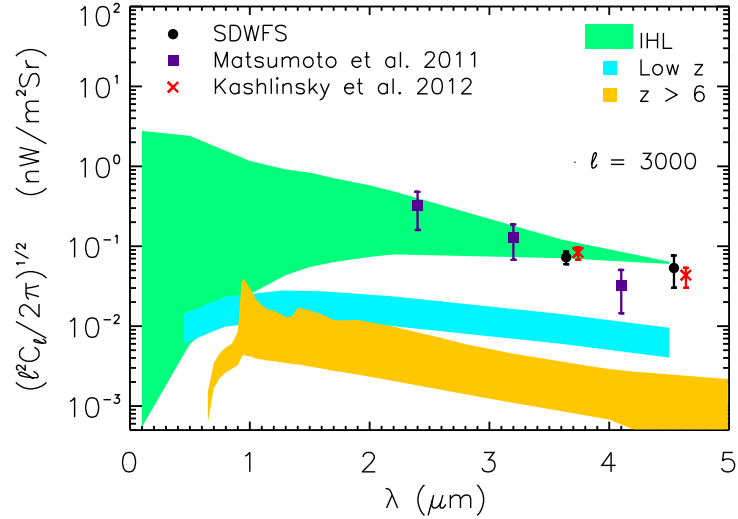}}
\caption{{\bf The spectral energy distribution of IR background anisotropies}. The frequency spectrum of near-IR and optical background anisotropies
as a function of the wavelength. We show the rms fluctuation amplitude at $\ell=3000$, corresponding to fluctuations at 6 arcminute angular scales.
We show our measurement and the existing measurements in the literature at the same angular scales\cite{Kashlinsky12,Matsumoto11}. 
The plotted error bars are the 1$\sigma$ uncertainties directly propagated to a rms fluctuation amplitude from the power spectrum errors.
We do not show the measurements at 1.6 and 1.1 $\mu$m \cite{Thompson08} as they
do not probe fluctuations on scales greater than 2 arcminutes on the sky due to the narrow field of view of the observations.
The shaded regions are (a) predictions for the IHL taking a variety of spectral energy distributions for the stripped stars (green),
(b) $z > 6$ galaxies (yellow), and (c) low-redshift galaxies below the detection level of masking (blue). 
}
\label{fig:sed}
\end{figure}

\clearpage



\setcounter{page}{1}

\setcounter{figure}{0}
\setcounter{table}{0}
\renewcommand{\figurename}{Figure~S}
\renewcommand{\tablename}{Table~S}

\begin{center}
{\bf \large Supplementary Information}
\end{center}

In these Supplemental Notes to the main paper, we outline key details related to how we
estimated the angular power spectrum  of {\it Spitzer}/IRAC data and its interpretation.
The data used for this study are publicly available from the Spitzer Heritage Archive\footnote{http://sha.ipac.caltech.edu/applications/Spitzer/SHA/}
under the program number GO 40839 (PI. D. Stern).

\section{The Spitzer Deep Wide Field Survey}

We use the IRAC data from the Spitzer Deep Wide Field Survey (SDWFS)\cite{Eisenhardt:2004,Ashby:2009}. Of the NDWFS Bo\"otes fields, the observations were obtained in 4 epochs,
with depth per pixel of 90 seconds in each epoch. Each epoch took observations over 7 to 10 days to complete the
full mosaic. We started with the basic calibrated data (BCDs).  
At each epoch, imaging data was obtained in all four IRAC wavelengths or channels (3.6, 4.5, 5.8, and 8 $\mu$m); however, for this study we focus on the  3.6 and 4.5 $\mu$m data.   Each epoch consists of
4300 to 4900 BCDs per channel that were mosaiced to form the final images used in the fluctuation analysis.

Instead of using the public SDWFS mosaics (Available at http://irsa.ipac.caltech.edu),
 made with the standard {\it Spitzer} data analysis pipeline MOPEX\cite{Makovoz05}, we produced our own
mosaics in order to better control systematic errors.   To mosaic the BCDs for each channel and epoch we used a self-calibration algorithm\cite{Fixsen:2000,Arendt} 
to properly match the sky background level from one adjacent frame to the other in the overlapping region 
using an optimized least squares fitting technique. The SDWFS mapping strategy incorporates several elements to facilitate
self-calibration of the data by maximizing inter-pixel correlations. We dithered the observations on small
scales and offset by one-third of an IRAC field-of-view between successive passes through each group. This provides inter-pixel
correlation information on both small and large scales, so the self-calibrated mosaic has background levels that are
stable across the wide area of the SDWFS mosaic. Finally, for the larger, rectangular groups, we cadence the observations
such that revisits cover the same area but with a different step
size.  With a $< 10$\% penalty in mapping efficiency, cadencing significantly 
enhanced the inter-pixel correlations across all scales.

These mapping strategies were designed to significantly enhance the
self-calibration of the data. Finally, by reobserving the field
multiple times at different roll angles, our observing strategy
was designed to be robust against bad rows/columns, large scale cosmetic
defects on the array, after-images resulting from saturation due
to bright stars, variations in the bias level, and the color
dependence of the IRAC flat-field across the array\cite{Quijada04}.
In particular, the challenging diffuse background measurements we report here
are vastly aided by the redundant coverage: independent data sets
of the same region are the best way to assess and control systematic errors. With
this mapping strategy we were able to construct independent sky
realizations for carrying out jackknife testing. Furthermore, the
power spectrum is estimated by cross-correlating the maps from different epochs,
eliminating bias from uncorrelated signals such as instrumental
noise and mosaicing artifacts. 

\begin{figure}[!ht]
    \begin{center}
      {\includegraphics[scale=0.6]{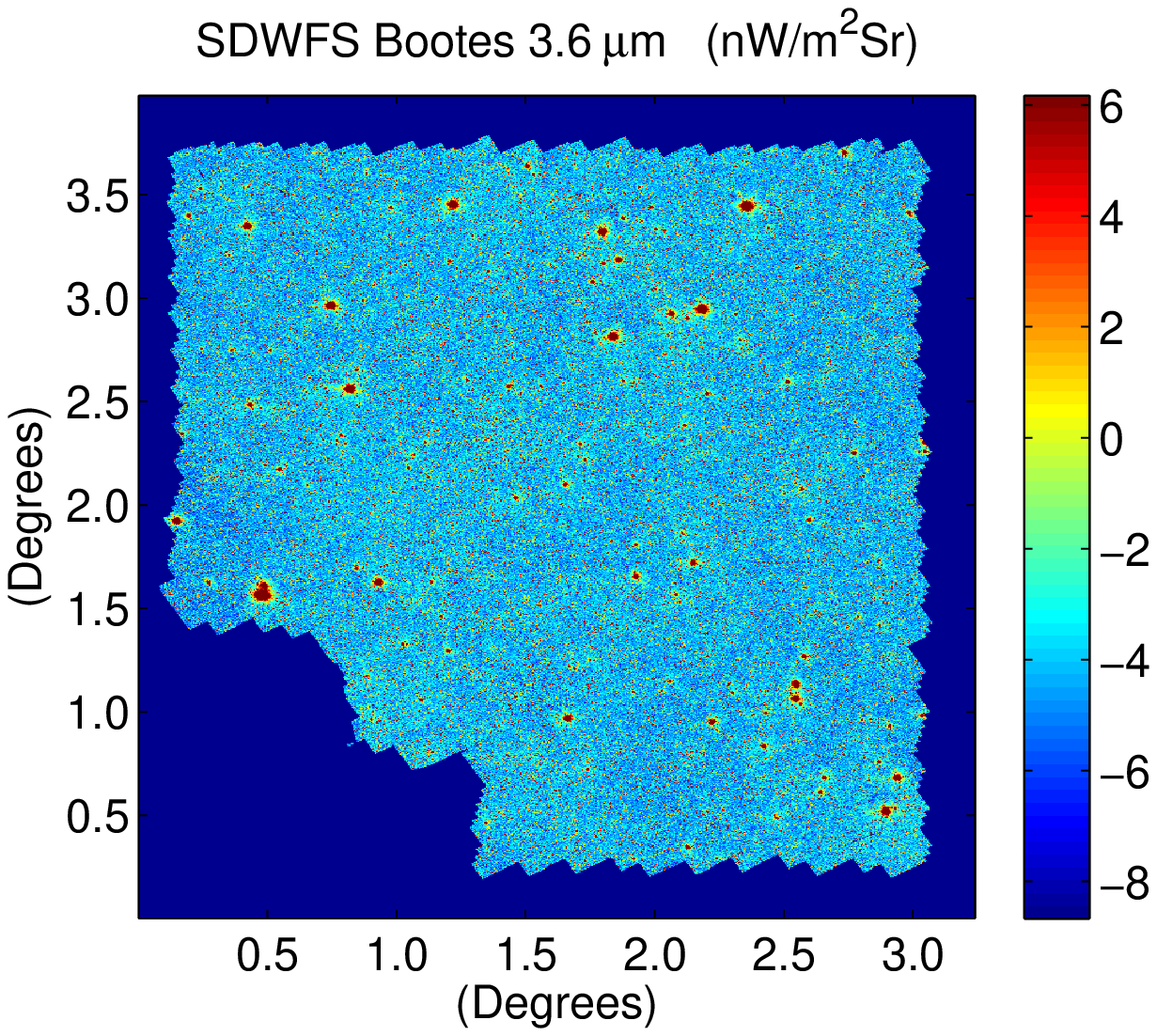},\includegraphics[scale=0.6]{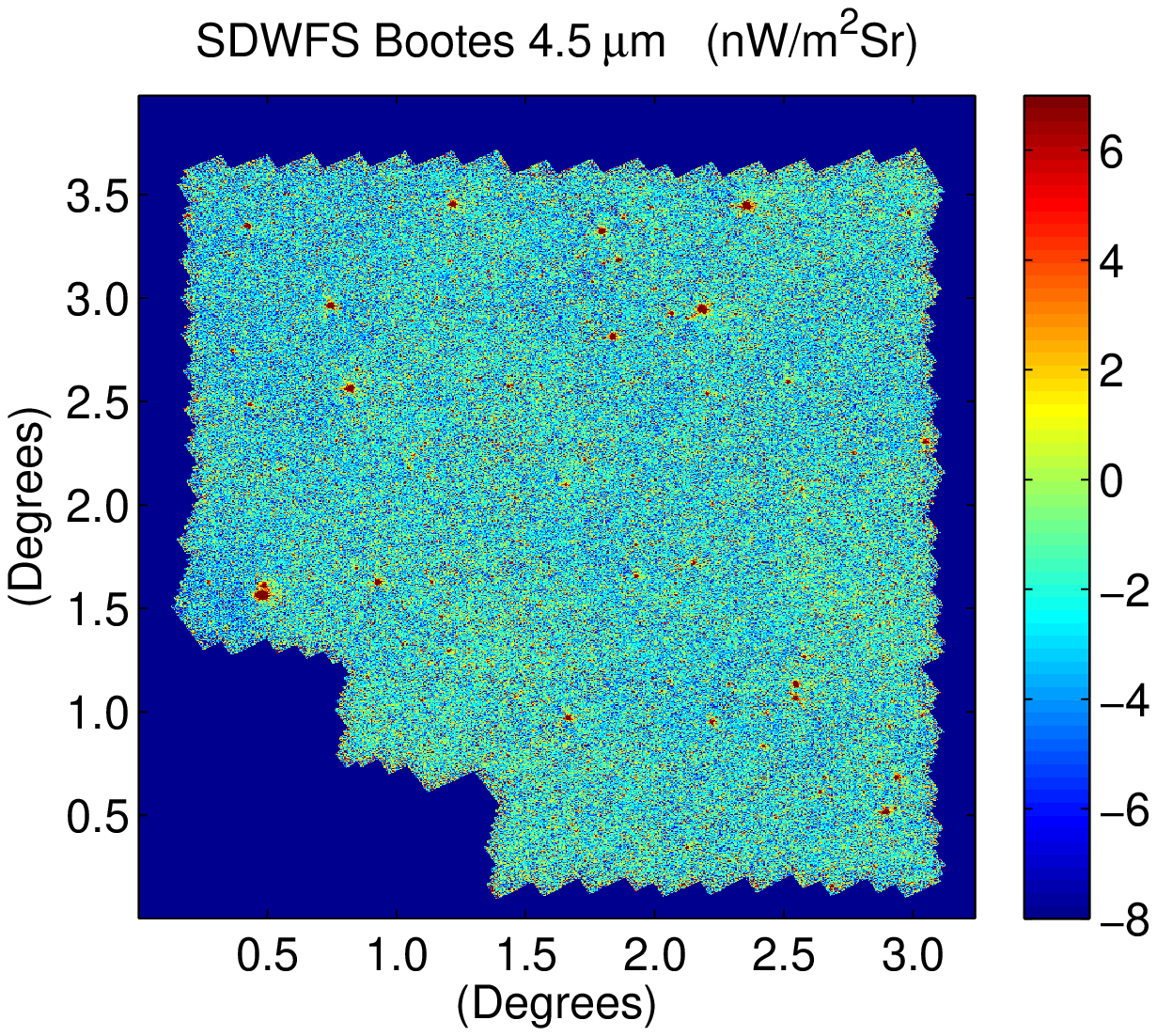}}
   \end{center}
   \caption{{\bf The SDWFS Maps. } The final 3.6 $\mu$m (left) and 4.5 $\mu$m (right) self-calibrated mosaics used for the Bo\"otes SDWFS analysis.  }
    \label{fig:sdwfs_map}
   \label{fig:kernel}
\end{figure}

For each IRAC channel and epoch we passed the cleaned BCD into our self-calibration code, a slightly modified version of an existing code\cite{Arendt},
 as inputs the cleaned BCDs (cBCDs) with the final, output  array size and astrometry defined to correspond
to the mosaic of first 3.6 $\mu$m epoch. The cBCDs were first cleaned of asteroid trails, hot pixels, and other image artifacts.
Since the astrometry is the same for each mosaic, they 
can be properly coadded and jack-knifed as described below.  Each of the cBCDs have an angular pixel scale of 1.2 arcseconds, which was preserved in the final mosaics.  The 
portion of the maps used for analysis are $\sim 3.5 \times 3$ degrees on a side for a total area of $\sim10.5$ square degrees. The final mosaics generated from the self-calibration 
algorithm are shown in Fig.~S\ref{fig:sdwfs_map}.

\section{Generation of the Detected Source Mask}

As the analysis concentrates on the background intensity fluctuations, with the aim of identifying the nature of faint sources below the individual detection level,
 we must remove the contamination from individual detected sources in our power spectrum measurement. We created source masks based both on the objects
detected in the IRAC data at 3.6 and 4.5 $\mu$m and the NOAO Deep Wide Field Survey (NDWFS) catalogs for the B$_{w}$, R and I bands.  
The masked sources include stellar point sources, galaxies that are extended, and galaxies that are unresolved but detected as point sources in SDWFS.
The mask also accounts for the Spitzer-IRAC point spread function (PSF). For this study we make use of the publicly available "extended" IRAC PSF to properly account for flux wings.

Next, we summarize our recipe to generate the source list and discuss more details below:
\begin{enumerate}
\item We create a catalog consisting of all objects detected by {\sc Sextractor} at a 3$\sigma$ detection threshold on the coadded
and individual epoch maps at both 3.6 and 4.5 $\mu$m.
\item We add any additional objects detected in the NDWFS optical catalogs, but missing from the IRAC catalogs.
\item We create a map initially of zeros where we place each source detected from our combined catalogs with the proper flux values. The sources that are flagged as extended in the SExtractor
catalogs are placed as extended sources with a size that is comparable to their estimated size.
\item We convolve this final map with the IRAC PSF to ensure that the flux wings for each source are properly masked.
\item We make a cut at a certain flux level so that all pixels with intensity down to that threshold are masked. We then histogram the remaining pixels from the data map
 and cut all pixels that are $\pm$ 5 $\sigma$ away in the histogram. The latter step is similar to prior approaches to measure the {\it Spitzer}-IRAC background anisotropy power spectrum\cite{Arendt10}.
\item To produce the final mask, we set the pixels with intensity values above the final flux and outliers from histogram to a value of zero, and the remaining pixels,
the ones used for anisotropy measurements, a value of one.
\end{enumerate}

Before the mosaics were created, source extractor was run iteratively on each epoch and waveband individually as well as on the coadded 
maps for both wavebands in order to find detected sources.  The parameters used for {\sc Sextractor} are the same as the ones used for the original SDWFS catalogs\cite{Ashby:2009}.
The combined catalog obtained from this iterative source extractor analysis, 
as well as the objects detected in the NDWFS catalog are merged into a final catalog.

\begin{figure}[!ht]
    \begin{center}
        {\includegraphics[scale=0.55]{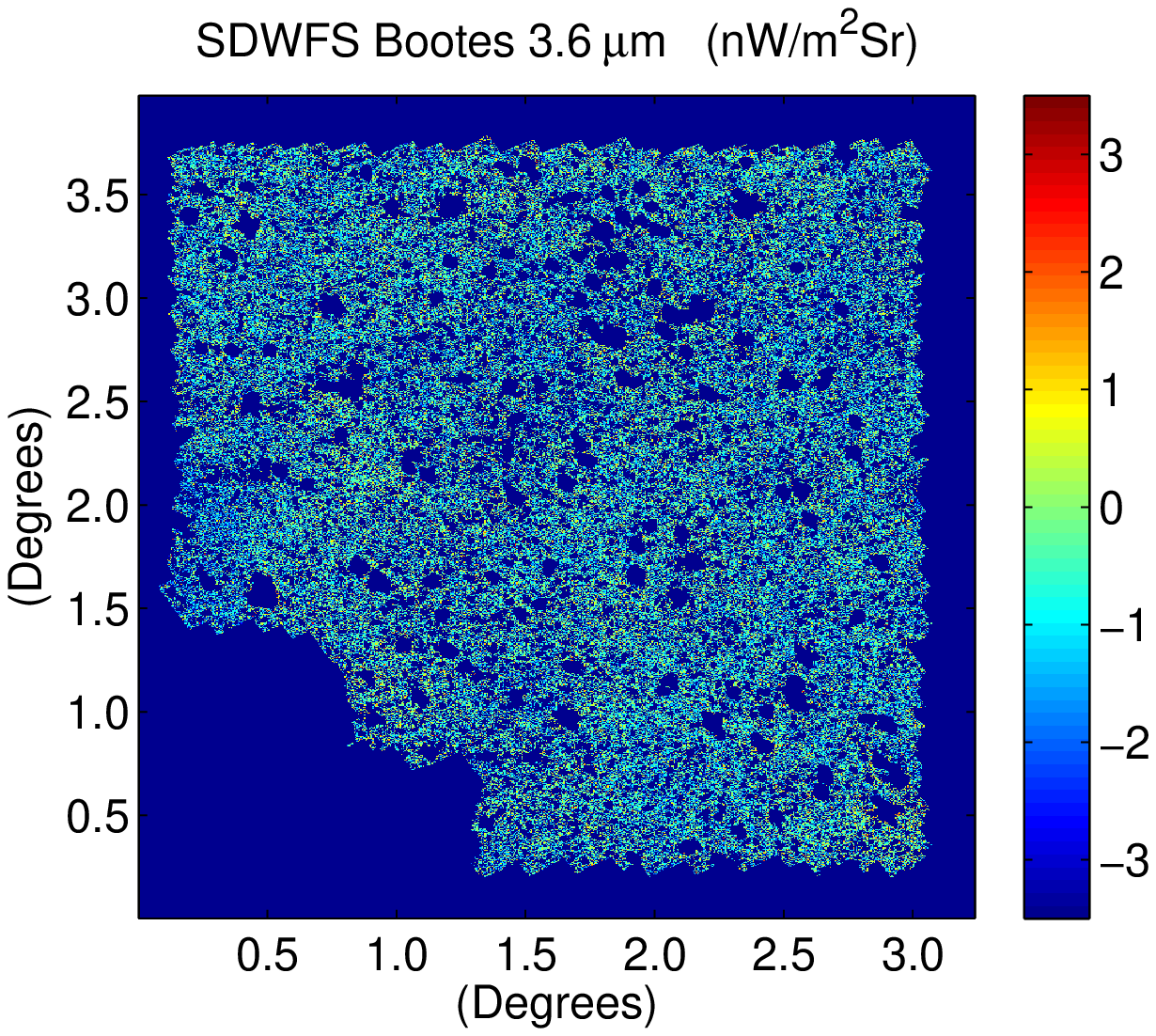},
        \includegraphics[scale=0.55]{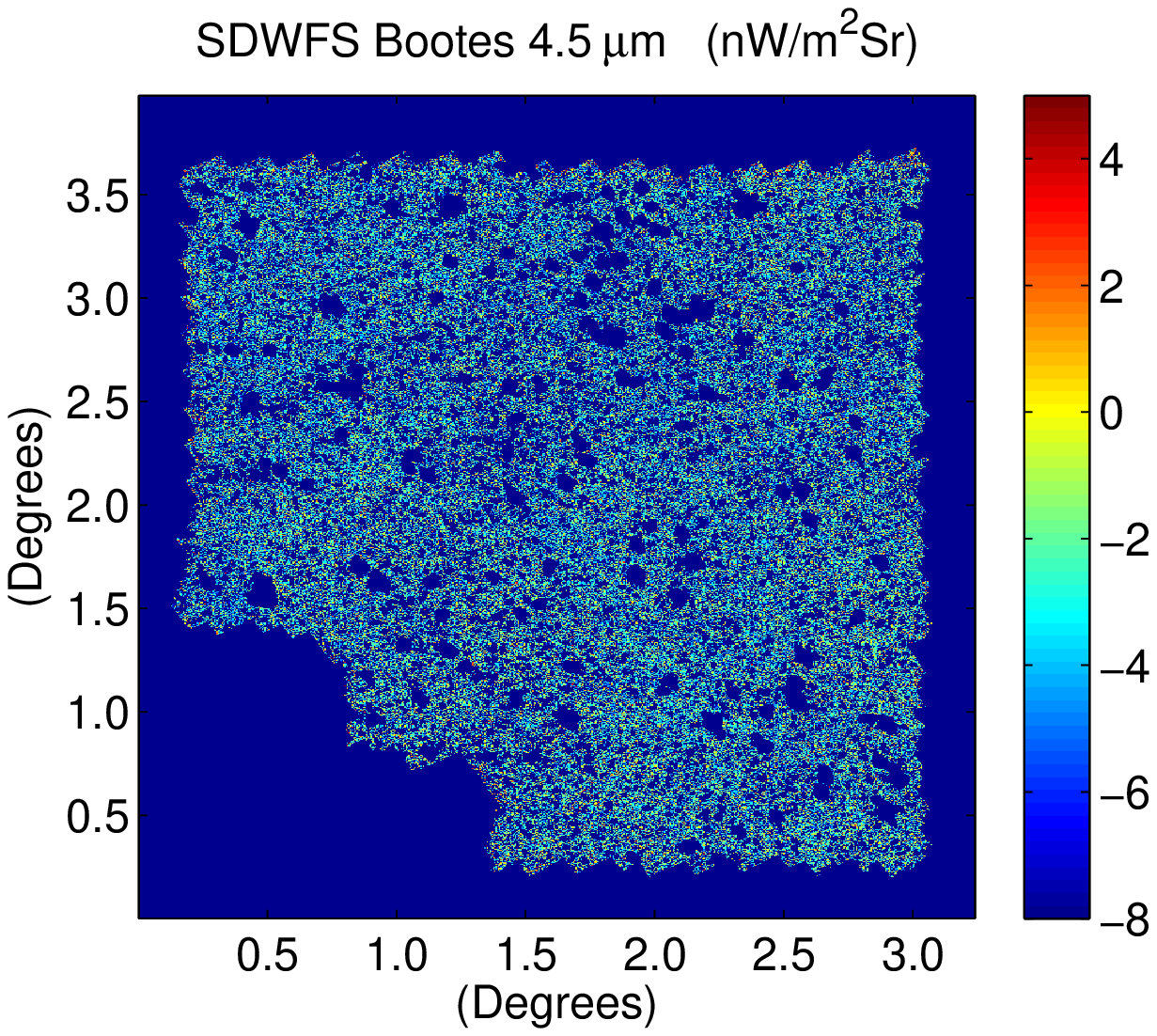}}
    \end{center}
    \caption{{\bf The Masked SDWFS maps.} The final 3.6 $\mu$m (left) and 4.5 $\mu$m (right) masked maps used for the Bo\"otes SDWFS analysis. The mask used for the analysis
removes 56\% of the pixels in the mosaic. }
    \label{fig:sdwfs_masked}
\end{figure}

The final flux cut for the mask was found iteratively by lowering the flux limit until further expansions of the masked regions no longer affect the final
power spectrum. Since the PSF is slightly different for the 3.6$\mu m$ and 4.5$\mu m$ wavebands the two masks are independently constructed.
Fig.~S\ref{fig:sdwfs_masked} shows the final masked epoch 1 maps. The final mask is such that 56\% of the original pixels are removed from the subsequent analysis.

\section{Power Spectrum Estimation}

With the final mosaics and masks in hand, initial, raw auto and cross-correlations may be computed to measure the level of clustering in the maps at each scale. To calculate the cross-correlation between masked maps $M_1$ and $M_2$ in real space, we first take the 2D Fourier transform of each map which we call
$\tilde{M}_1$ and $\tilde{M}_2$, as
\begin{equation}
\tilde{M}[l_x,l_y] = \Delta \sum_{m = 0}^{M-1}  \sum_{n = 0}^{N-1}  M[m,n] e^{-2\pi i (l_x m/M + l_y n/N  )}
\end{equation}
where M and N are the number of discrete points in the two dimensions of the map and $\Delta$ is the sampling interval in radians. 
The power spectra $C_l$ formed from the cross-correlation of $M_1$ and $M_2$ for a specific $l_i$ bin between between $l$-modes $l_1$ and $l_2$ is equal to the weighted mean of the squared Fourier modes $\tilde{M}_1\tilde{M}^*_2$ between $l_1$ and $l_2$,
\begin{equation}
C_{l_i} ={ \sum_{l_x^2 + l_y^2 \geq l_1}^{l_x^2 + l_y^2 \leq l_2} w[l_x, l_y] \tilde{M_1}[l_x,l_y]  \tilde{M_2^*}[l_x,l_y] \over \sum_{l_x^2 + l_y^2 \geq l_1}^{l_x^2 + l_y^2 \leq l_2} w[l_x, l_y] },
\end{equation}  
where $w[l_x, l_y]$ is a window function in Fourier space that is non-zero for each mode of the analysis and zero for modes that are discarded.
To compute the raw auto-correlation we have the special case where $M_1 = M_2$.

\begin{figure}[!ht]
    \begin{center}
        {\includegraphics[scale=0.7]{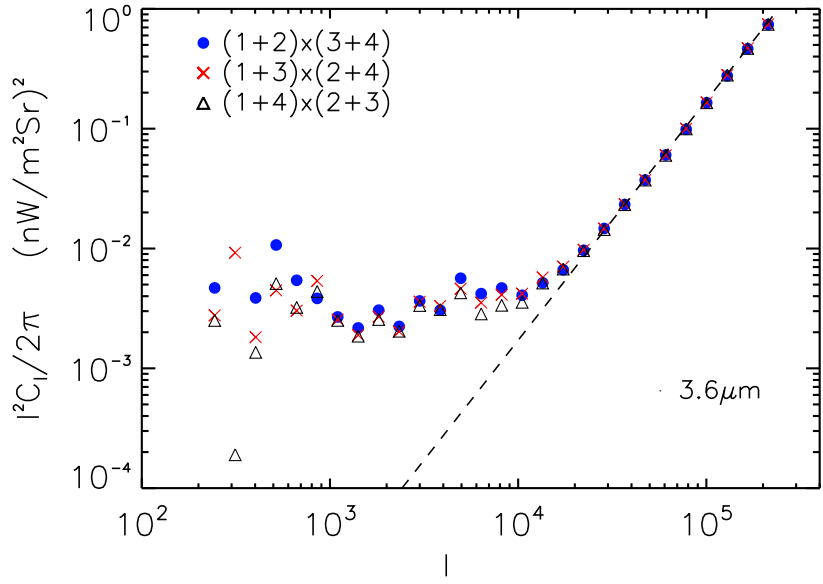},
        \includegraphics[scale=0.7]{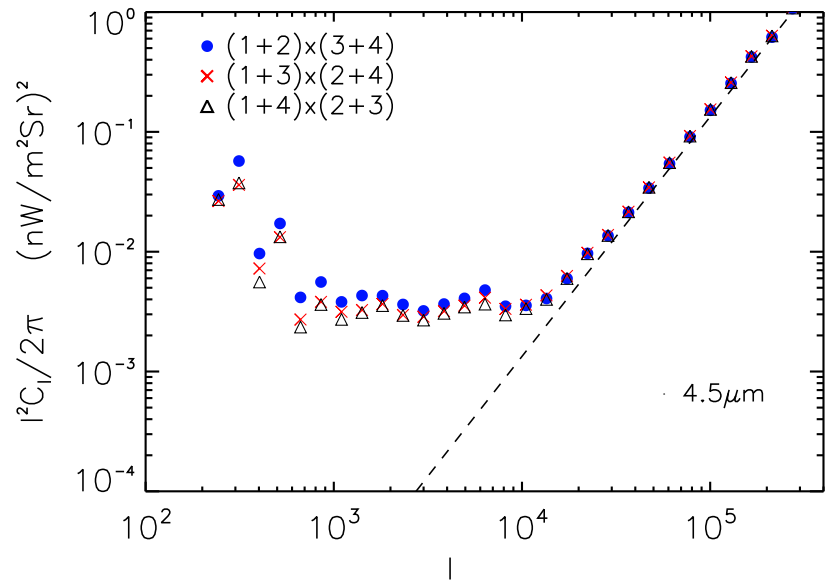}}
    \end{center}
\caption{{\bf The cross power spectra of the sum of multi-epoch maps}. 
The cross-correlation power spectra of different epoch summed maps with 3.6 $\mu$m (top) and 4.5 $\mu$m (bottom)
shown separately. The average of the summed maps are taken to be the power spectrum. 
The notation $(a + b) \times (c + d)$ indicates a cross correlation between the average of the $a + b$ and the $c + d$ epochs.}
   \label{fig:pow_all}
 \end{figure}

  \begin{figure}[!ht]
    \begin{center}
        {\includegraphics[scale=0.7]{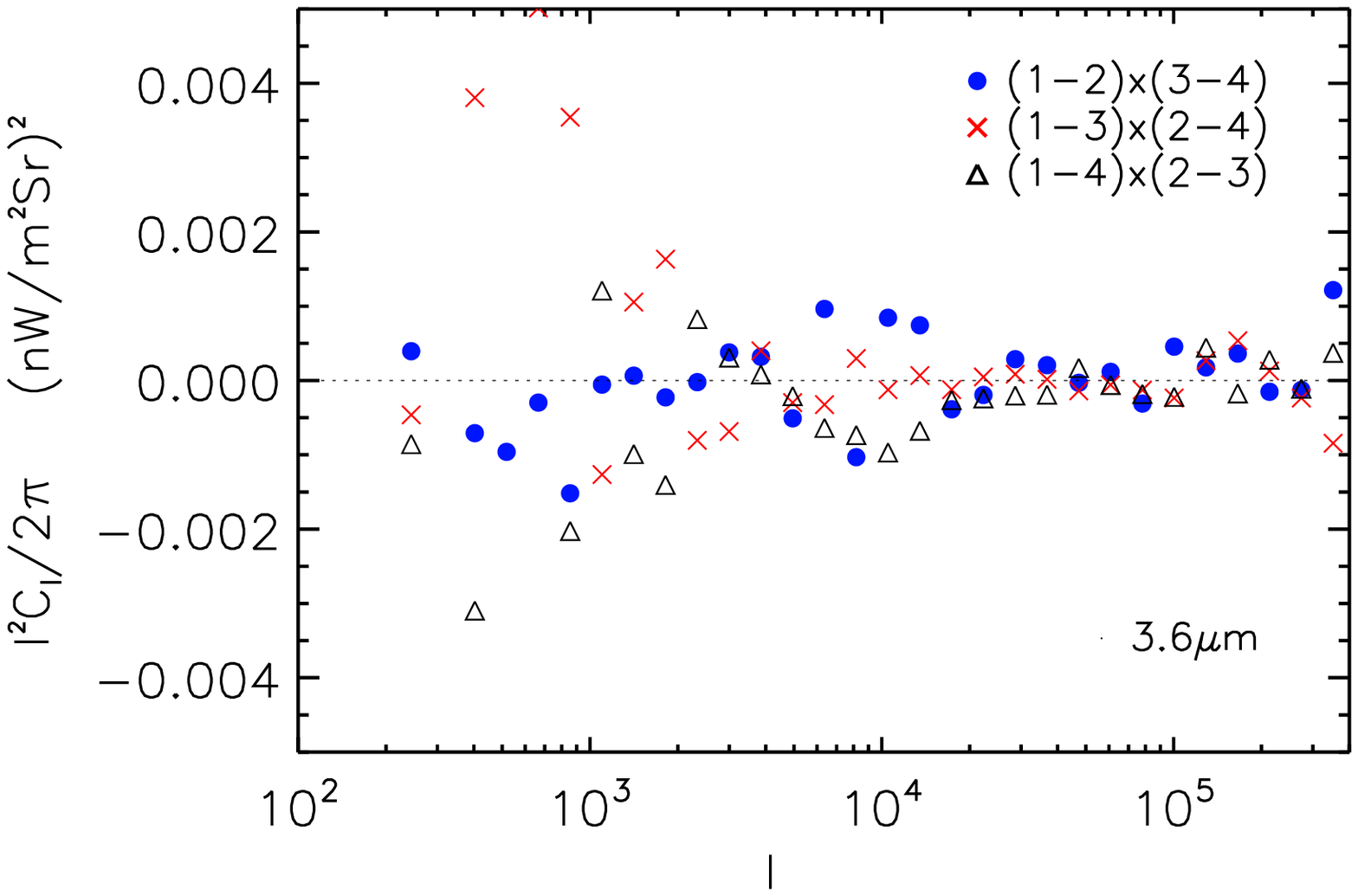},
        \includegraphics[scale=0.7]{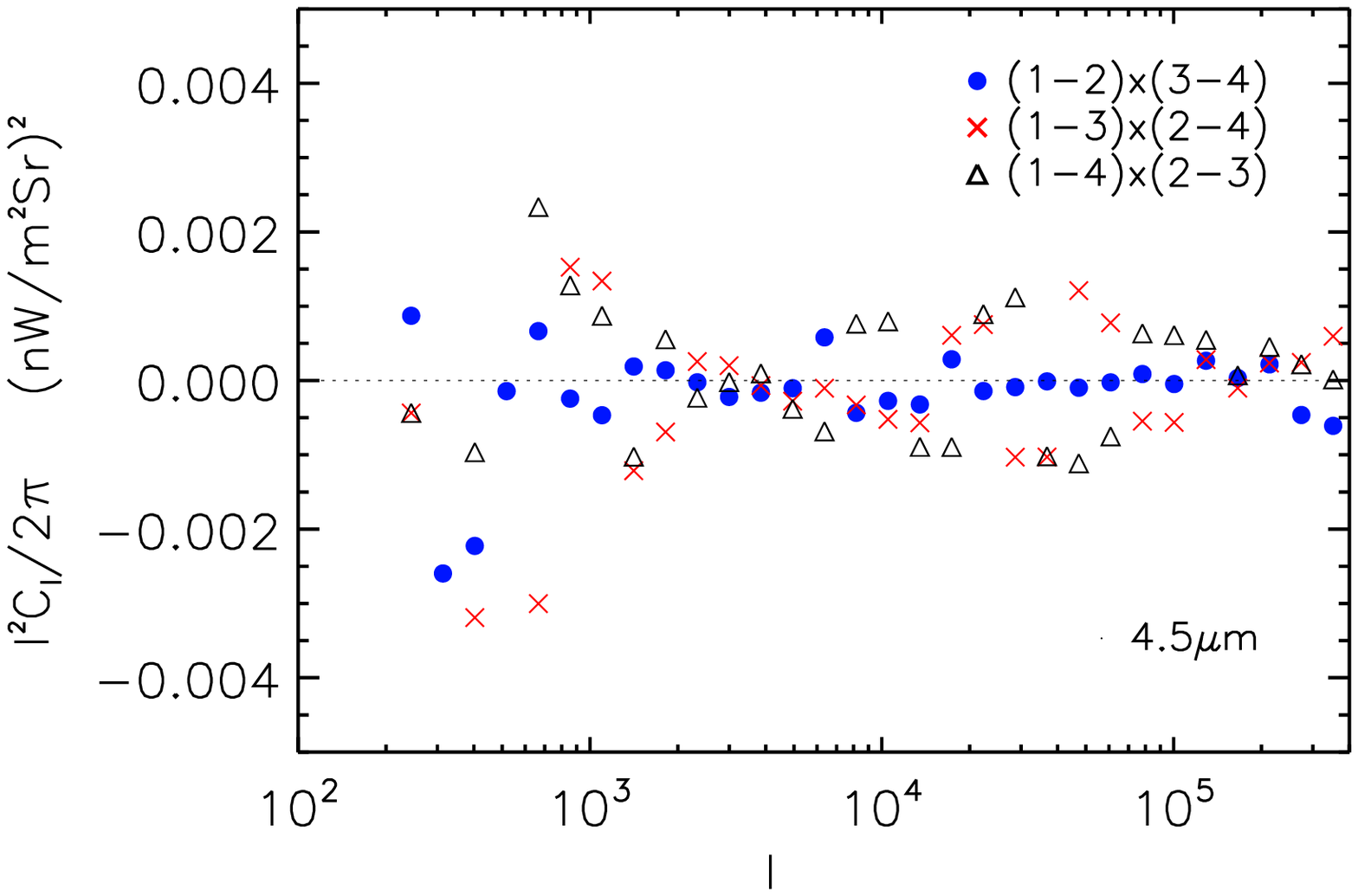}}
    \end{center}
    \caption{{\bf The cross power spectra of the difference of multi-epoch maps}. The cross-correlation power spectra of the difference of multi-epoch maps between epochs 1 to 4 with 3.6 $\mu$m (top) and 4.5 $\mu$m (bottom)
shown separately. The cross-correlations are consistent with zero
and the variance between the different cross-correlations provide  one part of the final error budget associated with the power spectrum measurement.}
   \label{fig:jacks}
 \end{figure}
 
   \begin{figure}[!ht]
    \begin{center}
        {\includegraphics[scale=0.7]{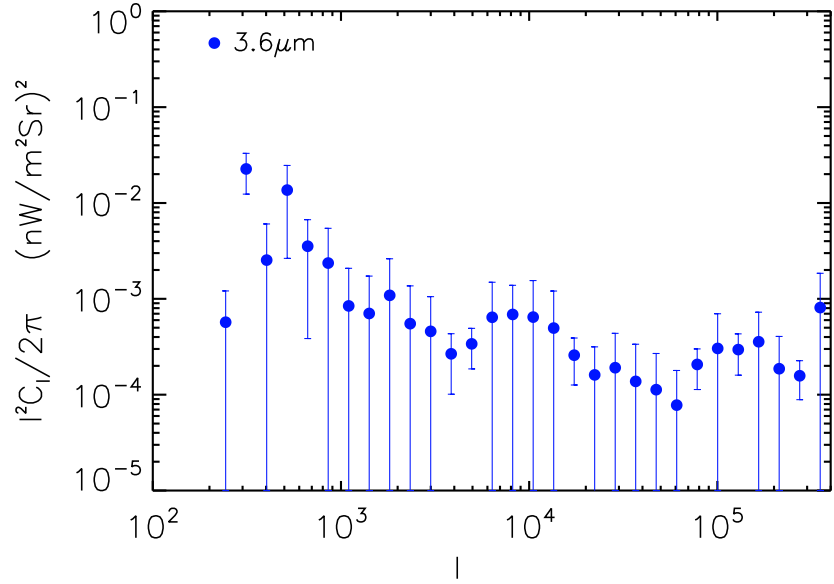},
        \includegraphics[scale=0.7]{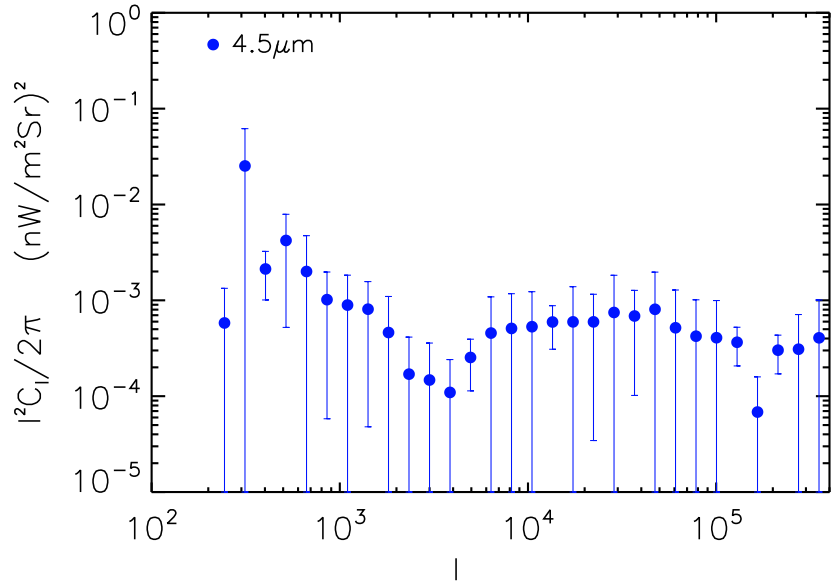}}
    \end{center}
    \caption{{\bf The noise floor to detect near-IR background anisotropies}. The mean and the variance ($1 \sigma$) 
of the different-epoch difference maps cross-correlations. We show the
absolute value of the mean and the error on the mean from the variance of the three independent cross-correlations.}
   \label{fig:jacksave}
 \end{figure}

 \begin{figure}[!ht]
    \begin{center}
        {\includegraphics[scale=0.6]{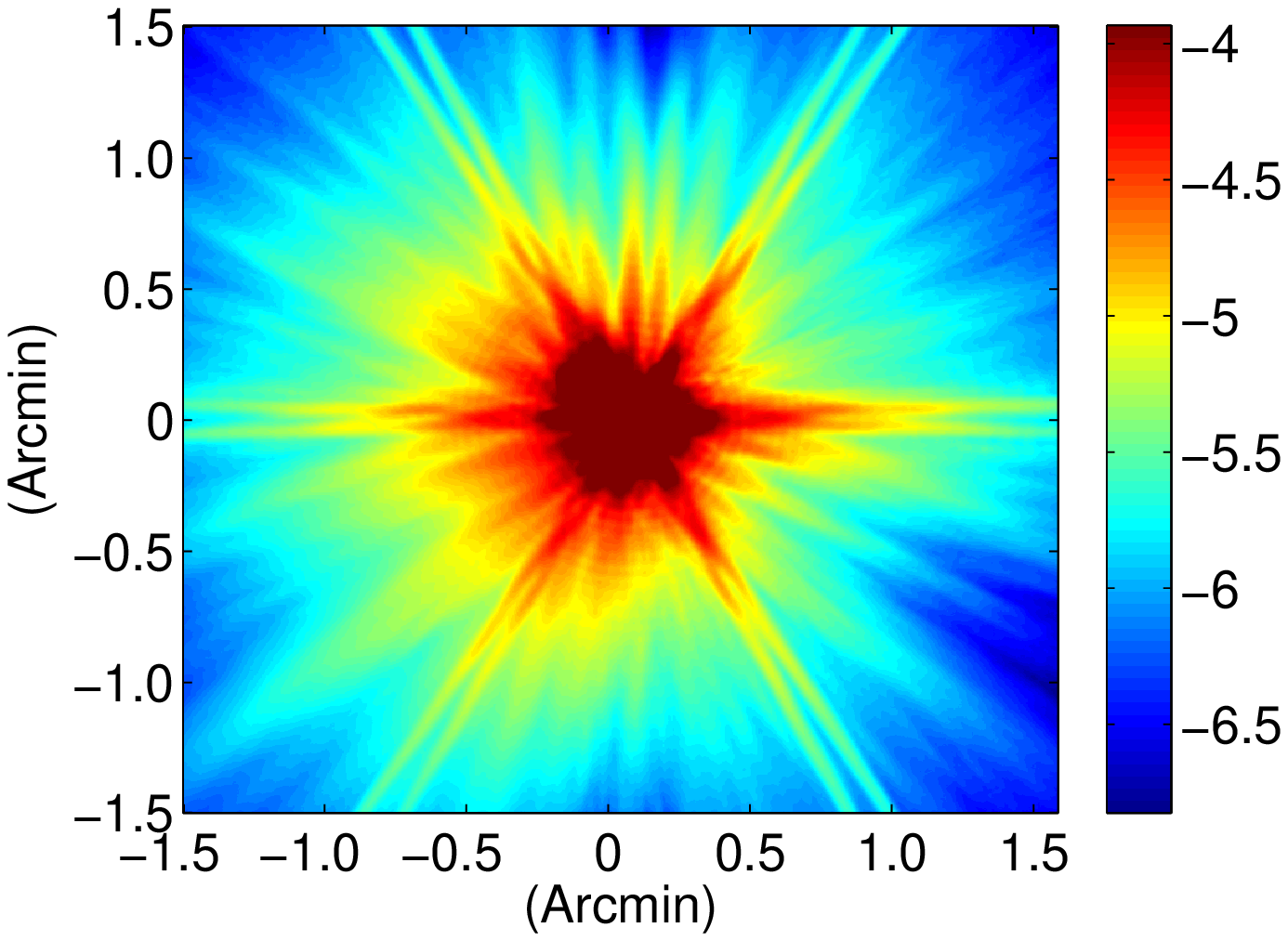},\includegraphics[scale=0.6]{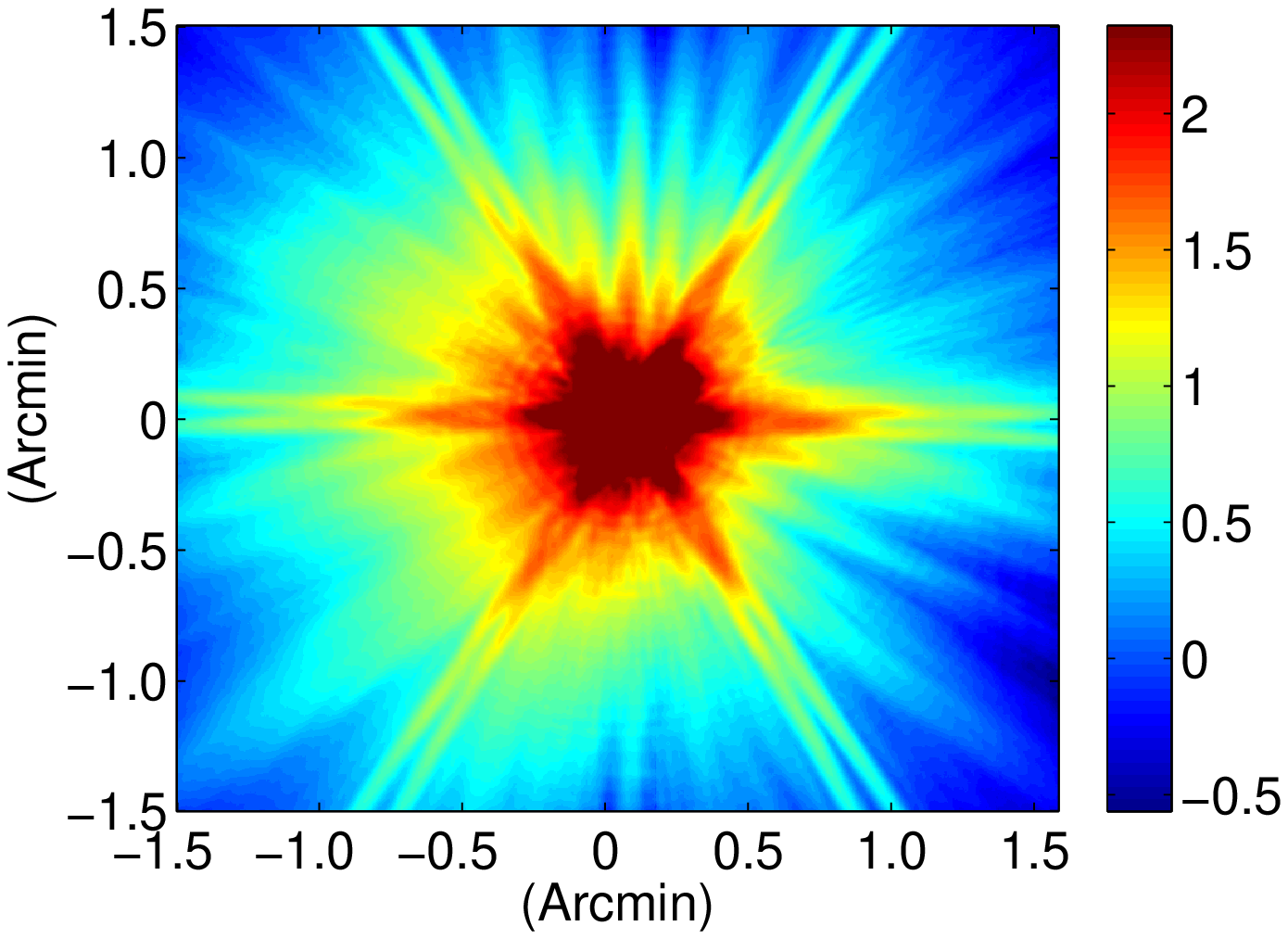}}
    \end{center}
    \caption{{\bf The IRAC PSF.} The 3.6 $\mu$m (left) and 4.5 $\mu$m (right) IRAC PSFs on a 
logarithmic intensity scale.}
   \label{fig:sdwfs_psf}
 \end{figure}

Being able to form a power spectrum from a cross-correlation rather than an auto-correlation is highly advantageous,
 as the noise bias and other contaminants that can dominate an auto-correlation calculation are minimized in a cross-correlation.  
This is because the pixel of each map $M_i=S_i+N_i$ is really a sum of the signal $S_i$ plus noise $N_i$. The noise contributes to the auto-correlation such that
$M_1 \times M_1= S_1^2 + N_1^2$, but is minimized in a cross-correlation  $M_1 \times M_2 = (S_1 + N_1) \times (S_2+N_2) = S_1^2$, since $S_1=S_2$.
Another advantage to a cross-correlation study in this analysis is the availability of multi-epoch data over a four-year period. In such data any time varying signals
that are not correlated between epochs cannot contribute to the background anisotropy power spectrum. One such possibility is the zodiacal light associated with scattered Sunlight
off of dust particles. This is due to dynamical dust particles in near-Earth orbits. Furthermore with varying {\it Spitzer} orbit the lines of sights to the SDWFS Bo\"otes
field  during the four epochs will also be different. Thus, we expect the zodiacal light contamination to have a time varying component
that is not correlated between epochs. While autocorrelations in single epochs may be contaminated by zodiacal light, we expect cross-correlations to reduce the 
contamination. A previous analysis\cite{Arendt10} showed that the contamination from zodiacal light spatial fluctuations is at least an order of magnitude
below the background anisotropy level at 3.6 $\mu$m. Thus, zodiacal light should not be the dominant systematic effect in the present 
analysis. The cross-correlations using sum maps of epochs 1 to 4 are shown in Fig.~S\ref{fig:pow_all}.

We can check the assumption that $N_1\times N_2$ terms cancel by examining the cross-correlation between, say, $M_1-M_2=N_1-N_2$ and $M_3-M_4=N_3-N_4$. For differences
of the same region the signal terms cancel and the amplitude of the cross-correlation $(M_1-M_2)\times(M_3-M_4)$ provides an estimate for the floor level at which noise contributions cancel.
In Fig.~S\ref{fig:jacks} we show these differences, where we find that the cross-correlation power spectra between different epoch differences are consistent with zero. The variance of these
difference cross-correlations provide a part of the error budget associated with noise correlations between different epochs including any systematic effects that
are not canceled out in difference maps and are correlated between epochs. Since these spectra represent the noise floor the absolute value, we add them to the 
final error budget in quadrature with other uncertainties (Fig.~S\ref{fig:jacksave}).

Even after cross-correlation, the raw spectra are contaminated by several different sources and require additional corrections.
These issues include resolution damping from the beam, the fictitious correlations introduced by the mask and shot noise.  All three of these contaminants are dealt with as described below.

\begin{figure}[!ht]
    \begin{center}
        {\includegraphics[scale=0.7]{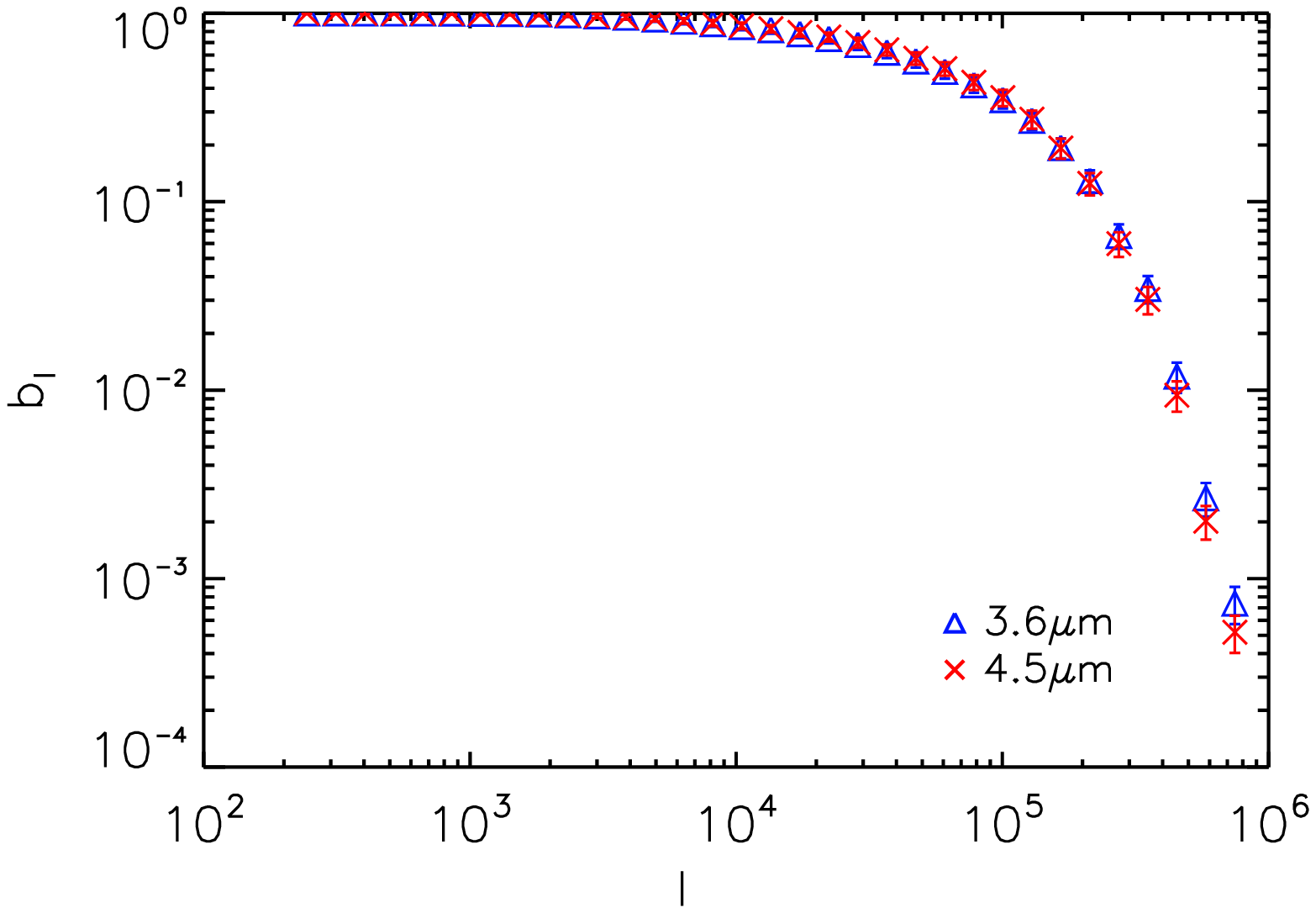}}
    \end{center}
    \caption{{\bf The IRAC beam function.} The beam transfer functions for the SDWFS Bo\"otes field in the multipole space $\ell$. We show the functions at 3.6 
and 4.5 $\mu$m with triangles and crosses, respectively. The plotted error bars are the 1 $\sigma$ uncertainties.}
    \label{fig:boot_beam}
\end{figure}

\section{The Beam Correction}

The first correction that needs to be applied to the raw power spectra is the correction for the beam.  Realistic detectors have limits to their resolving power which causes a fictitious drop in power at high multipoles. 
For a known beam structure, the resolution limits of the instrument can be modeled
in harmonic space with a function that encodes the full-width-half-max (FWHM) of the telescope.  The resulting scale dependent function is known as the beam transfer function, $b_l$. 
For a Gaussian beam it can be computed analytically as
\begin{eqnarray}
b_l &=& \exp(l^2 \sigma_{\rm beam}^2/2) \hspace{1cm} and \\
\sigma_{\rm beam} &=& {\theta_{\rm FWHM} \over \sqrt{8 \ln{2}}},
\end{eqnarray}
where $\theta_{\rm FWHM}$ encodes the full-width-half-max of the
instrument's resolving power.  The Spitzer team has measured $\theta_{\rm FWHM} = 1.9$ arcsec. The beam transfer function can also be computed directly from the data  by measuring the point spread function.
Each bright point source in the sky really shines a thin point like beam of light that should only illuminate a single pixel of the detector.  However, due to 
the finite resolution  of the telescope, the source is spread out over many pixels and often has a complex shape very different from a Gaussian. 
Fig.~S\ref{fig:sdwfs_psf} shows this is true of the SDWFS PSF, and for
this reason the beam transfer function needs to be calculated directly from the PSF.

In general, the beam transfer function $b_l$ in Fourier or multipolar space is
\begin{equation}
\label{eq:uds_bl}
b_l^2 = {C_l^{M_{\rm psf}} \over C_l^{M_{\rm point}}}
\end{equation}
where $C_l^{M_{\rm psf}}$ is the power spectrum of $M_{{\rm psf}}$, the observed image of a point source including the
effects of the telescope, and $C_l^{M_{\rm point}}$ is the power spectrum of a true point source where all the light lies in one pixel of the map. Fig.~S\ref{fig:boot_beam} shows the beam calculated using Eq.~\ref{eq:uds_bl} and the publicly available models of the {\it Spitzer}-IRAC  PSF.
The PSFs differ for each epoch and waveband and the appropriate beam transfer function for
a cross-correlation between maps $M_1$ and $M_2$ becomes:
\begin{equation}
b_l^{1 \times 2} = \sqrt{b_l^1 b_l^2}
\end{equation}
where $b_l^1$ and $b_l^2$ are the beam transfer functions for maps $M_1$ and $M_2$ respectively.

The beam corrected spectra $C_l$ are then computed from the raw power spectrum $C_l^{raw}$ by dividing by the beam transfer function $b_l$,
\begin{equation}
C_l = C_l^{raw}/b_l^2.
\end{equation}
To measure the uncertainty in the beam transfer function we must understand the uncertainties in the PSF.
The SDWFS team provides several measured PSFs taken across the Bo\"otes images from which the uncertainty in $b_l$ can be measured from Eq.~\ref{eq:uds_bl} 
as $\delta b_l = {\delta C_l^{M_{\rm psf}} / C_l^{M_{\rm point}}}$ with $\delta C_l^{M_{\rm psf}}$ estimated from
the variance of the differences from using 
the various PSF models. 

    \begin{figure}
    \begin{center}
        {\includegraphics[scale=0.55]{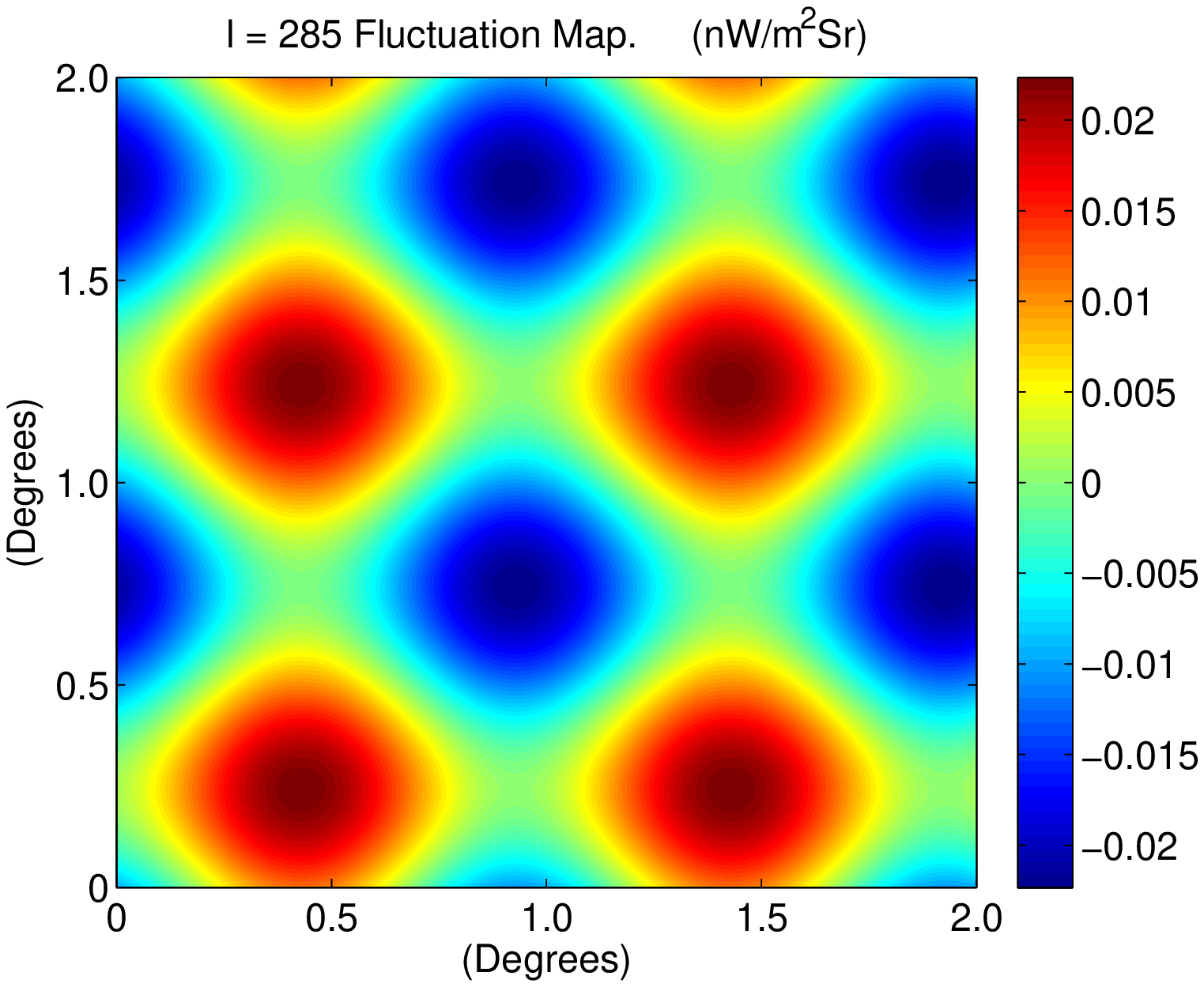},\includegraphics[scale=0.55]{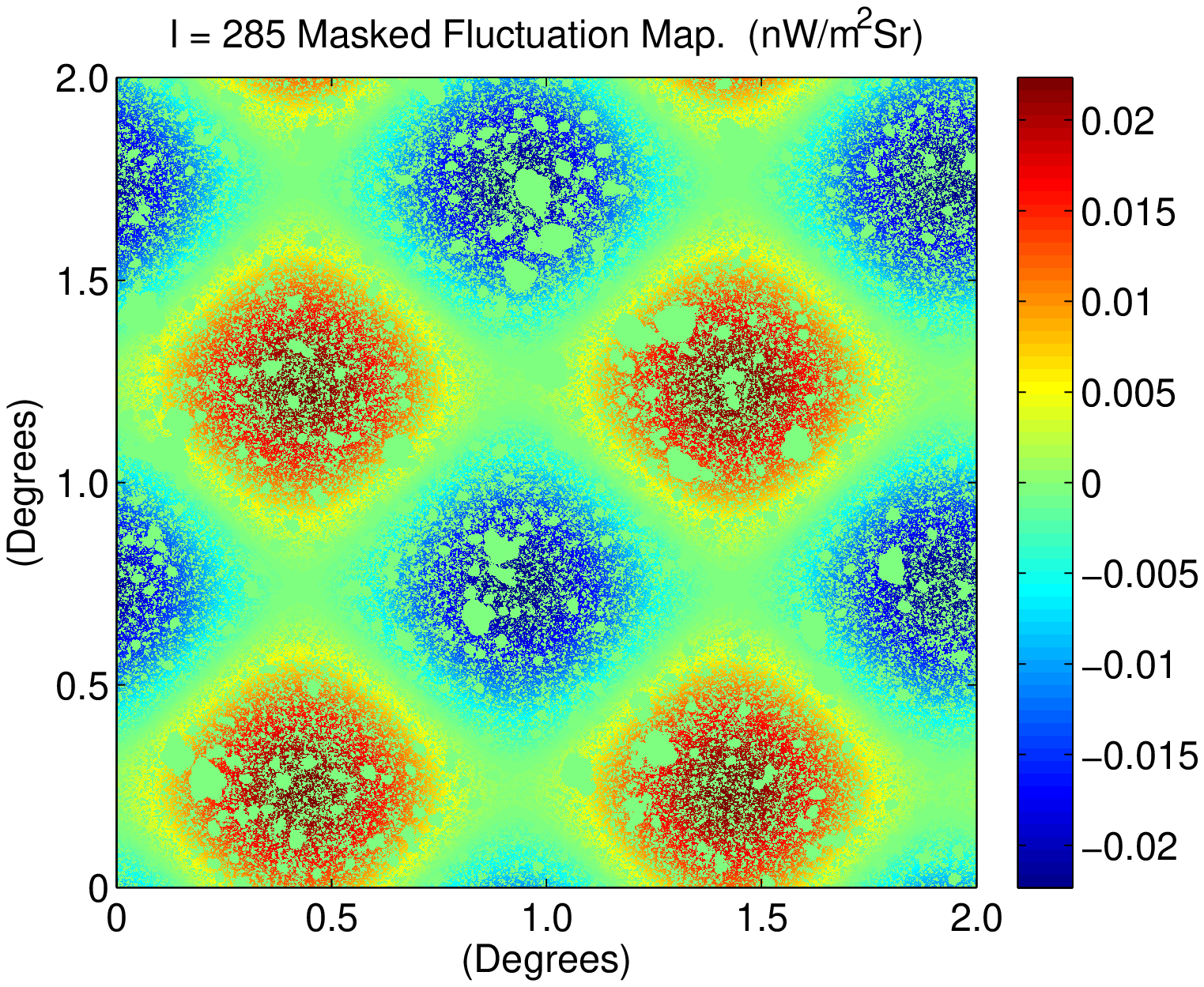}}
    \end{center}
    \caption{{\bf The masking effects in a map.} The left shows a map of a $l = 285$ fluctuation.  
On the right is the same map masked by one of the SDWFS masks.  
Note how what was only a large scale fluctuation gets broken up into smaller modes by the mask, contaminating the true power.}
   \label{fig:flucs_sim}
 \end{figure}

     \begin{figure}
    \begin{center}
        {\includegraphics[scale=0.7]{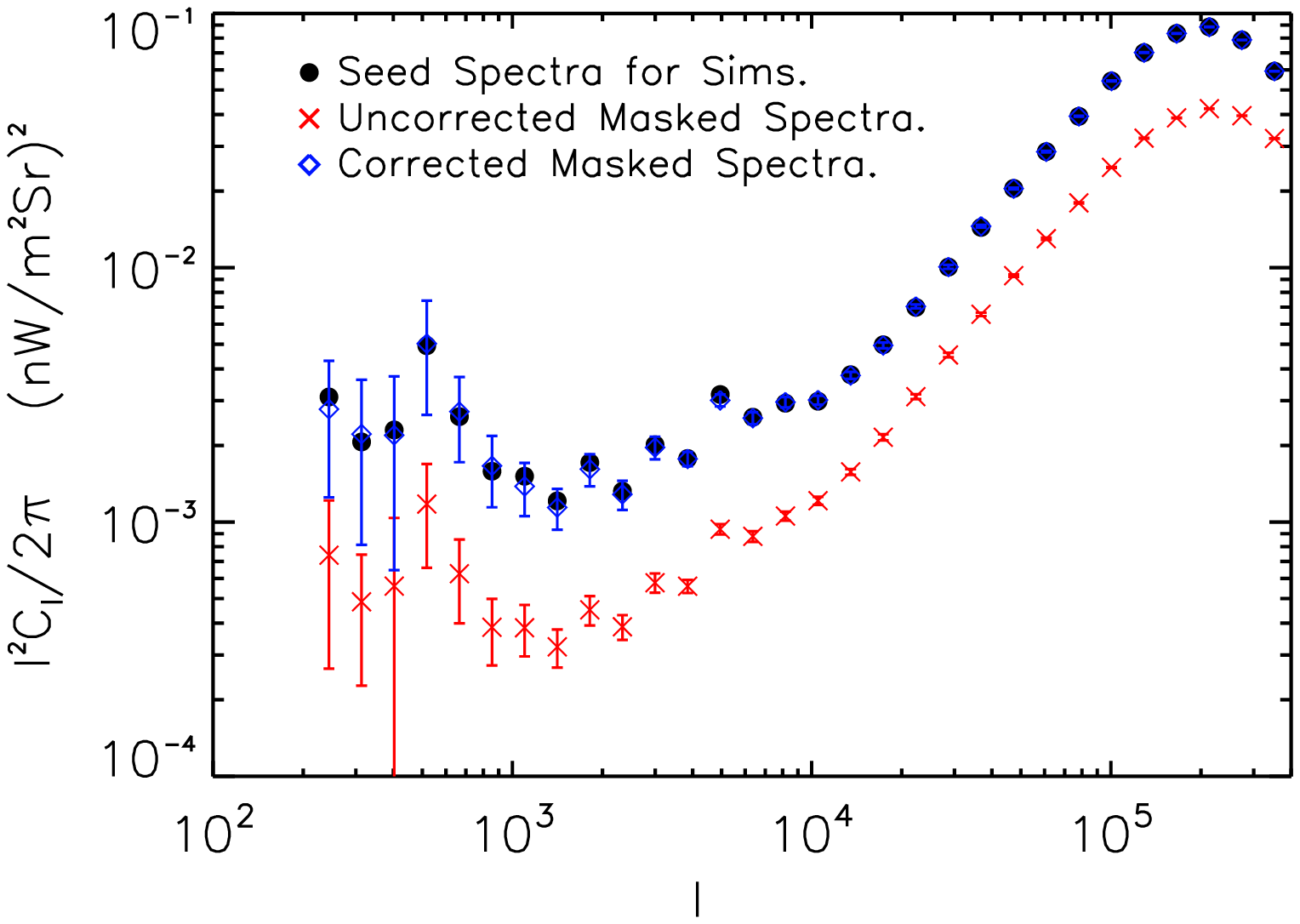},
        \includegraphics[scale=0.7]{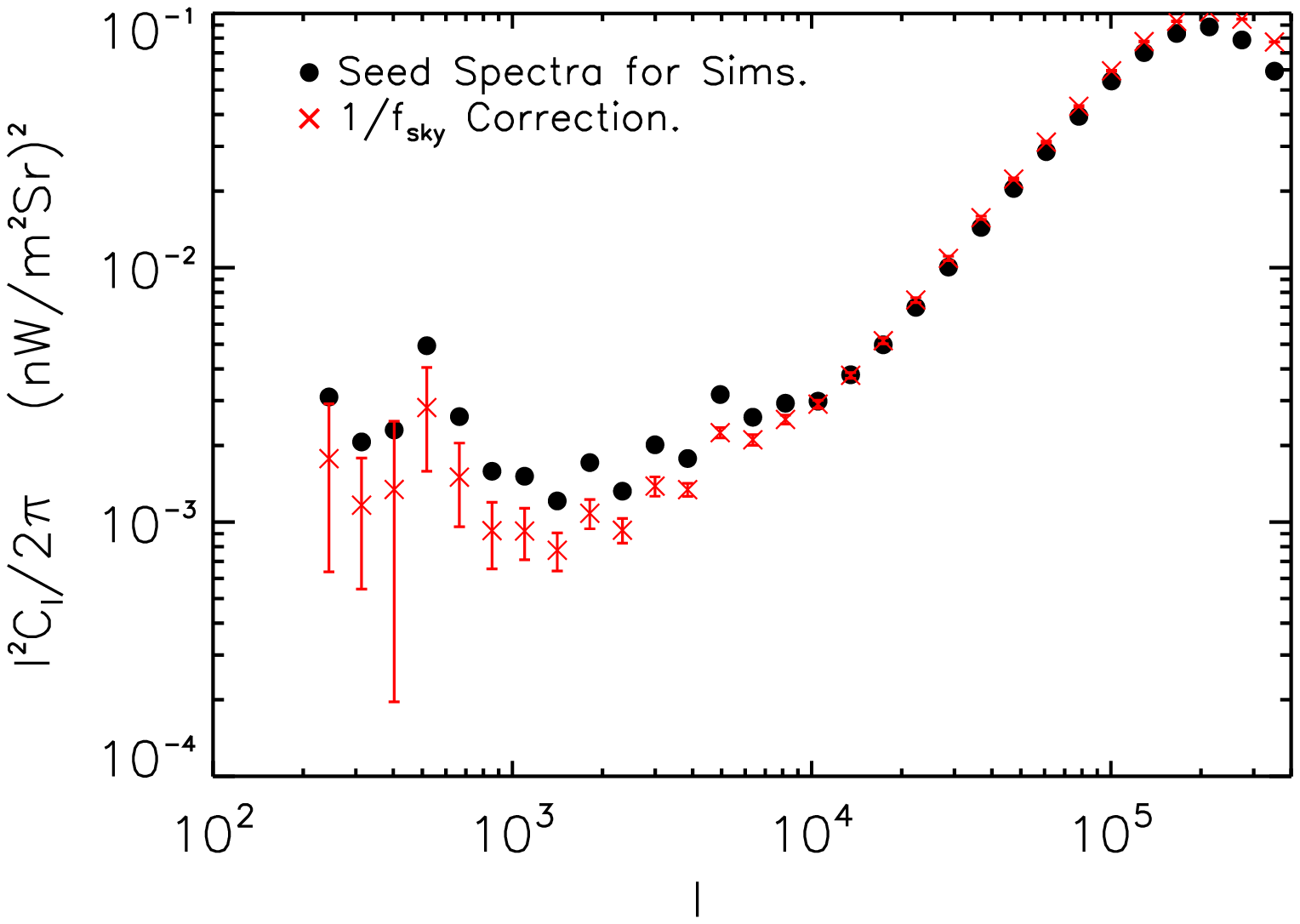}}
    \end{center}
    \caption{{\bf Masking effects on the power spectrum.} 
The top panel shows how well we can recover the true spectrum of a masked sky using the mode coupling matrix. The black points show the seeded spectra we used to create 100 simulations.  The red shows what happens to the spectra after masking using the $3.6$ $\mu$m mask.  The blue points
show what happens after correcting with the mode-coupling matrix.  
The bottom panel  shows the result if corrected with only the masked sky fraction given by $f_{\rm sky}$. The mask breaks many large modes into smaller modes, 
so that after the $f_{sky}$ correction the large-scale modes are under-represented and the small-scale modes are over represented.  
This illustrates the need to use the full coupling matrix to correct for the mask.
The plotted error bars are the 1 $\sigma$ uncertainties.}
   \label{fig:mkk_sim}
 \end{figure}

\begin{table}
\begin{center}
  \begin{tabular}{c|c|c}
  \hline
  $l_{\rm eff}$ & $l^2 C_l /2\pi$ (3.6 $\mu$m) [nW m$^{-2}$ sr$^{-1}$]$^2$ & $l^2 C_l /2\pi$ (4.5 $\mu$m) [nW m$^{-2}$ sr$^{-1}$]$^2$\\
    \hline
$243$ & $ (0.27 \pm 0.36) \times 10^{-2} $ &  $ (2.04 \pm 2.82) \times 10^{-2} $ \\
$313$ & $ (0.52 \pm 0.58) \times 10^{-2} $ &  $ (2.81 \pm 3.33) \times 10^{-2} $ \\
$402$ & $ (0.18 \pm 0.24) \times 10^{-2} $ &  $ (0.54 \pm 0.49) \times 10^{-2} $ \\
$517$ & $ (0.51 \pm 0.50) \times 10^{-2} $ &  $(0.89 \pm 0.87) \times 10^{-2} $  \\
$665$ & $ (0.32 \pm 0.26) \times 10^{-2} $ & $ (0.23 \pm 0.23) \times 10^{-2} $ \\
$854$ & $ (0.43 \pm 0.42) \times 10^{-2} $ &  $( 0.28 \pm 0.18) \times 10^{-2} $\\
$1099$ & $ (0.25 \pm 0.12) \times 10^{-2} $ &  $ (0.23 \pm 0.12) \times 10^{-2} $\\
$1412$ & $ (0.18 \pm 0.11) \times 10^{-2} $ &  $ (0.24 \pm 0.12) \times 10^{-2} $\\
$1815$ & $ (0.26 \pm 0.16) \times 10^{-2} $ & $ (0.27 \pm 0.08) \times 10^{-2} $ \\
$2332$ & $ (0.19 \pm 0.09) \times 10^{-2} $ & $ (0.21 \pm 0.04) \times 10^{-2} $ \\
$2997$ & $ (0.34 \pm 0.08) \times 10^{-2} $ & $ (0.20 \pm 0.03) \times 10^{-2} $ \\
$3851$ & $ (0.29 \pm 0.05) \times 10^{-2} $ &  $ (0.22 \pm 0.04) \times 10^{-2} $ \\
$4949$ & $ (0.43 \pm 0.09) \times 10^{-2} $ &  $ (0.24 \pm 0.05) \times 10^{-2} $ \\
$6360$ & $ (0.32 \pm 0.13) \times 10^{-2} $ & $ (0.28 \pm 0.09) \times 10^{-2} $ \\
$8173$ & $ (0.36 \pm 0.13) \times 10^{-2} $ &  $ (0.22 \pm 0.08) \times 10^{-2} $ \\
$1.05\times 10^4$ & $ (0.34 \pm 0.10) \times 10^{-2} $ &  $ (0.22 \pm 0.09) \times 10^{-2} $ \\
$1.35\times 10^4$ & $ (0.42 \pm 0.08) \times 10^{-2} $ & $ (0.25 \pm 0.10) \times 10^{-2} $ \\
$1.735\times 10^4$ & $ (0.53 \pm 0.05) \times 10^{-2} $ & $ (0.35 \pm 0.11) \times 10^{-2} $ \\
$2.229\times 10^4$ & $ (0.72 \pm 0.03) \times 10^{-2} $ &  $ (0.54 \pm 0.12) \times 10^{-2} $ \\
$2.865\times 10^4$ & $ (1.02 \pm 0.04) \times 10^{-2} $ &  $ (0.71 \pm 0.16) \times 10^{-2} $ \\
$3.682\times 10^4$ & $ (1.49 \pm 0.04) \times 10^{-2} $ &  $ (1.03 \pm 0.17) \times 10^{-2} $ \\
$4.731\times 10^4$ & $ (2.14 \pm 0.03) \times 10^{-2} $ & $ (1.48 \pm 0.19) \times 10^{-2} $  \\
$6.081\times 10^4$ & $ (3.05 \pm 0.03) \times 10^{-2} $ &  $ (2.10 \pm 0.15) \times 10^{-2} $ \\
$7.814\times 10^4$ & $ (4.28 \pm 0.05) \times 10^{-2} $ &  $ (2.97 \pm 0.13) \times 10^{-2} $ \\
$1.004\times 10^5$ & $ (5.87 \pm 0.06) \times 10^{-2} $ & $ (4.11 \pm 0.14) \times 10^{-2} $  \\
$1.291\times 10^5$ & $ (7.67 \pm 0.09) \times 10^{-2} $ &  $ (5.27 \pm 0.12) \times 10^{-2} $ \\
$1.658\times 10^5$ & $ (8.99 \pm 0.08) \times 10^{-2} $ & $ (6.15 \pm 0.06) \times 10^{-2} $ \\
$2.131\times 10^5$ & $ (9.28 \pm 0.04) \times 10^{-2} $ &  $ (5.92 \pm 0.12) \times 10^{-2} $ \\
$2.739\times 10^5$ & $ (7.67 \pm 0.02) \times 10^{-2} $ &  $ (4.84 \pm 0.08) \times 10^{-2} $ \\
$3.52\times 10^5$ & $ (5.21 \pm 0.14) \times 10^{-2} $ &  $ (3.06 \pm 0.07) \times 10^{-2} $ \\
$4.523\times 10^5$ & $ (3.25 \pm 0.16) \times 10^{-2} $ &  $ (1.70 \pm 0.15) \times 10^{-2} $ \\
    \hline
  \end{tabular}
\end{center}
  \caption{ The final SDWFS power spectrum values $l^2 C_l /2 \pi$ for both the 3.6 $\mu$m and 4.5 $\mu$m bands. The quoted error is the 1 $\sigma$ uncertainty of the final
power spectrum.}
  \label{tab:cl_vals}
\end{table}

\section{The Mode Coupling Correction.}

Fictitious correlations introduced by the mask must be corrected.  When an image is masked, the sources are replaced by the value zero in the image.  When the power spectrum in computed, these zeros in real space make fictitious contributions to the two-dimensional Fourier transform that are then added to the final power spectrum.  This can be easily seen in Fig.~S\ref{fig:mkk_sim}.  On the top we see an unmasked fluctuation  pattern for a specific $l$-mode.  After the mask is applied, this fluctuation gets broken up into fluctuations of different sizes causing both a diminishing and a reshuffling of power in Fourier space.

A matrix correction method exists\cite{Hivon01} to model the effects of the mask on the power spectrum by using a matrix $M_{ll'}$ whose inverse removes the effects of the mask from the measurement by matrix multiplication.  
If $\tilde{C}_l$ is the masked sky power spectrum and $C_l$ be the true power spectrum, the relation between the  true and masked sky spectrum is
\begin{equation}
\label{eq:mkk_corr}
\tilde{C}_l = M_{ll'} C_{l'}\ , ,
\end{equation}
where Einstein summation notation is being used.
Since this relationship is matrix multiplication, the masking effects
can be removed from the masked power spectrum by simply using a matrix
multiplication  $M_{ll'}^{-1}\tilde{C}_l$ to recover
the true power spectrum $C_l$.  
In the limit of no $l-$mode coupling, $M_{ll'} =f_{\rm sky}$ where $f_{\rm sky}$ is the fraction of the masked map that is non-zero.

Calculating the mode coupling matrix $M_{ll'}$ analytically\cite{Hivon01} is computationally expensive for large maps.
For this reason, we developed a new way to generate the mode-coupling matrix as follows:
\begin{enumerate}
\item For each $\ell$ in the power spectrum create many realizations of
maps consisting of a pure tone where $C_m=1$
if $\ell=m$ and $C_m=0$ otherwise (an example case is shown in
Fig.~S\ref{fig:mkk_sim}).
\item For each of these trial maps, mask the maps and calculate an observed power spectrum $\tilde C_m(\ell)$.
\item The mode coupling matrix $M_{\ell m}$ = $\left<\tilde{C}_m(\ell)\right>$ is the average of the masked power spectra
found for the random realizations of model $\ell$. The inverse of the mode coupling matrix gives the sky power spectrum
corrected for the masking effect, $C_l = M_{l m}^{-1} \tilde C_m$.
\end{enumerate}

To see how well this this works, consider Fig.~S\ref{fig:mkk_sim}.  The black line shows the exact power spectrum 
from which we drew 100 simulated images.
The red points show the observed power spectra of the masked sky.
The blue points show those same 100 power spectra after correcting with the mode coupling matrix described above.  
This mode-coupling transformation does an excellent job recovering the
input power spectra of an unmasked sky.
As the lower panel of Fig.~S\ref{fig:mkk_sim} shows using the simplified model based only on  the masked sky fraction $f_{\rm sky}$ is not a good approximation.

To estimate the effects of cosmic variance on the power spectrum one
must make multiple Gaussian simulations with the exact power spectrum measured in the data.  
Without applying a mask, these simulations will have a cosmic variance:
\begin{equation}
\label{eq:cos_var}
\delta C_l^{\rm CV} = \sqrt{{2 \over (2l+1) \delta l f_{\rm sky}}} C_l
\end{equation}
where $\delta C_l^{\rm CV}$ is the cosmic variance for multipole $l$, $\delta l$ is the width of that $l$ bin,
$f_{\rm sky}$ is the fraction of the total sky covered by the unmasked region,
and $l$ in the denominator is the mid-point of the $l$ bin.
It should be clear that
masks should increase cosmic variance because they further reduce
the coverage of the unmasked sky.  The mode-coupling matrix does not
restore errors from cosmic variance and therefore, masking and
correcting with $(M_{ll'})^{-1}$ will combine errors from the mode-coupling matrix as well as the reduced coverage from the mask and thus have a greater variance then cosmic variance alone.

\begin{figure}
    \begin{center}
        \includegraphics[scale=0.6]{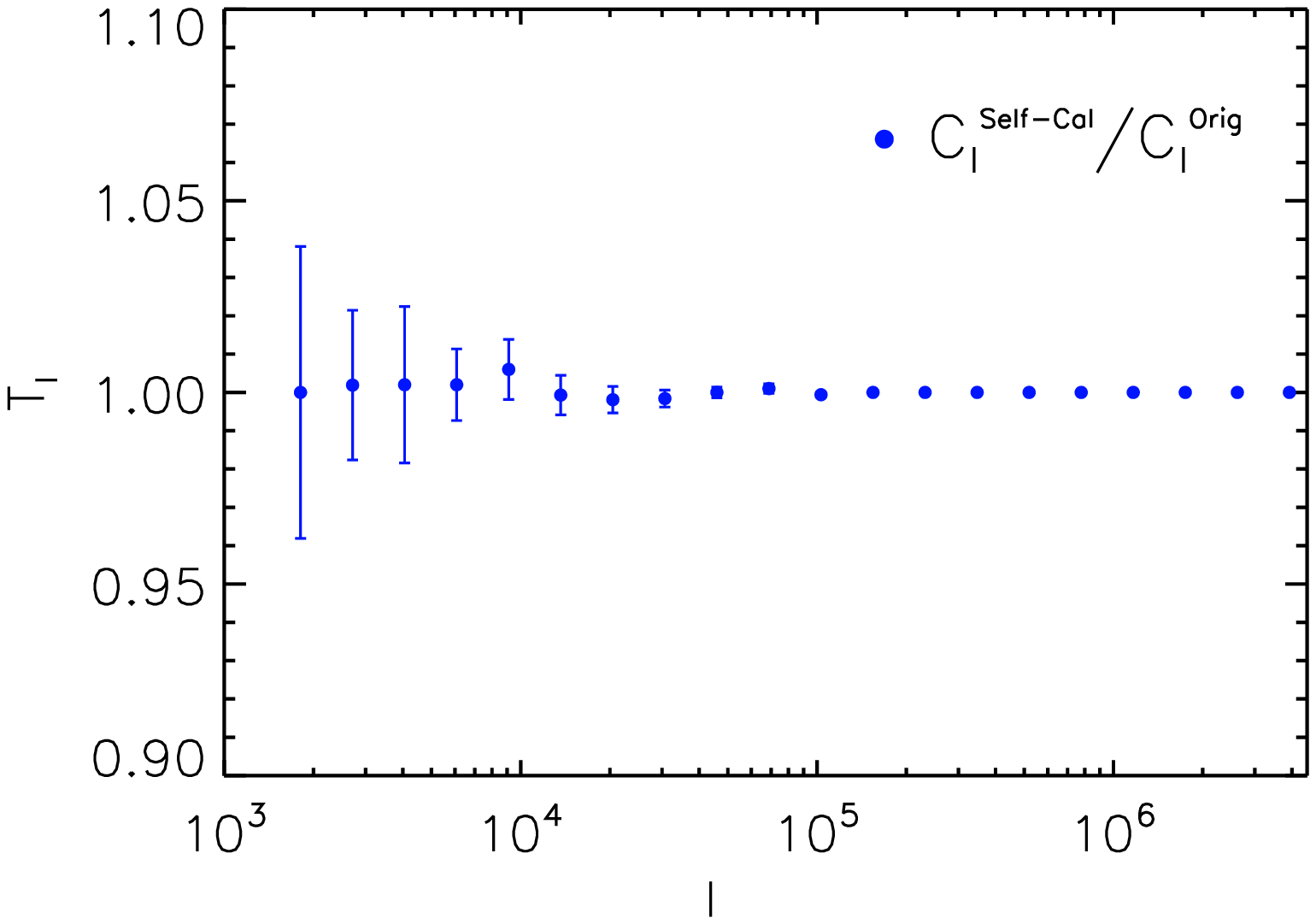}
    \end{center}
    \caption{{\bf The map-making transfer function.} {\it Spitzer}-IRAC mosaic transfer function based on the self-calibration algorithm used to make data maps for this study. We show
the transfer function at 3.6 $\mu$m. The 4.5 $\mu$m transfer function is similar to the result at 3.6 $\mu$m.
The plotted error bars are the 1 $\sigma$ uncertainties.}
   \label{fig:sdwfs_trans}
 \end{figure}

\section{The Map-making Transfer Function}

The last correction that must be accounted for is to correct for the effects of the map-making procedure in 
producing the final mosaic.
The mosaicing procedure is often combines many images and between dealing with overlap regions as well adding more tiles to a mosaic,
 the addition or subtraction of total power must be quantified and corrected.
The best way to uncover the effects of the mosaicing procedure on the
power spectrum is to take a map  with a known power spectrum, re-build
the map using your mosaicing procedure and compare the output and input
power spectra.  More specifically to build the map-making transfer function we followed the following procedure:
\begin{enumerate}
\item We created a large map similar in size, pixel scale and astrometry to our SDWFS data mosaic, but assuming a shape for the power spectrum (we started with pure white noise for which $C_l =$ 
constant).
\item We break the map into the tiles similar in size, pixel scale and
astrometry to our original data tiles. We added instrumental noise to
the tiles consistent with the noise of the different epochs.
\item We ran these simulated signal plus noise tiles through the same self-calibration mosaicing procedure described in Section~1 to produce 
a new map identical to the original map modulo the mosaicing effects.
\item  The map-making transfer function is then $T_l$ = $C_l^{\rm
orig}/C_l^{\rm mosiaced}$, where $C_l^{\rm orig}$ is the known power
spectrum of the original simulation and $C_l^{\rm mosiaced}$ is the power of the final map.
\item We repeated Steps 1 to 4 above for different initial power spectra to test if the transfer function remains the same or is different. We found that the
transfer function is independent of the assumption for the input power spectrum shape or the amplitude.
\end{enumerate}

For our analysis we used this process with the self-calibration algorithm for  the mosaic transfer function.
Simulated maps of pure white noise broken into tiles and remosaiced 
using the self-calibration algorithm to determine the transfer functions.  An example transfer function for SDWFS is given in Fig.~S\ref{fig:sdwfs_trans}.
This was obtained by simulating 10$^3$ independent maps and using the mean and the standard deviation of $(T_l)_i$, where $i$ denotes each simulation,
to determine the best-determined transfer function and its error.

Given the transfer function $T_l$, beam correction $b_l$, the
mode-coupling matrix $M_{ll'}$ the final power spectrum is estimated as
\begin{equation}
C_l = M^{-1}_{ll'} T_l' \tilde{C}_{l'}/b_{l'}^2 
\label{eq:uds_final}
\end{equation}
where $\tilde{C}_{l'}$ is the raw power spectrum after masking out the 
foreground sources and $C_l$ is the final corrected power spectrum.  
$C_l$ is the power spectrum that we present in the main paper and compared to previous results.

\begin{table}
\begin{center}
\begin{tabular}{lr}
\hline
$A_f$                                                                                                                                 & $0.0015\pm0.0002$\\
$\log(M_{\rm min}/M_{\odot})$                                                                                           &       $9.03\pm0.05$\\
$\log(M_{\rm max}/M_{\odot})$                                                                                        &       $11.91\pm0.05$\\
$\beta$                                                                                                                         &    $0.094\pm0.005$\\
$\alpha$                                                                                                                       &       $1.23\pm0.09$\\
$C^{\rm SN}_\ell$ (nW$^2$ m$^{-4}$ sr$^{-1}$)                                                                                          &  $(9.8\pm0.5)\times10^{-11} $\\
\hline
\end{tabular}\label{parameters}
\end{center}
  \caption{The best-fit parameter values of the IHL anisotropy power spectrum model to the 3.6 $\mu$m data using MCMC model fits. 
The quoted error bars are the 1$\sigma$ uncertainties
for each of the parameter likelihoods  marginalizing over other parameters.
}
  \label{tab:fit}
\end{table}%

\section{Final Power Spectrum Results}

To compute the power spectrum for the infrared background at  3.6 $\mu$m 
we used all four epochs of the  SDWFS data as described above.  
In order to minimize contamination from instrumental noise, zodiacal light and other systematics we used the
cross-correlation $1/2(E_1+E_2)\times1/2(E_3+E_4)$ and the two
additional permutations by switching the epochs, 
where $E_1..E_4$ correspond to the four epochs.
Masks were generated to remove the foreground 
and applied as described above before the cross-correlations were taken.
We used the same cross-correlation procedure to obtain the  4.5 $\mu$m power spectrum.
After the raw spectra are obtained from the mosaics, they are corrected for the beam, mode-coupling, mosaicing (or map-making) effects 
as described above using Equation~\ref{eq:uds_final}.

\begin{figure}
    \begin{center}
        {\includegraphics[scale=0.7]{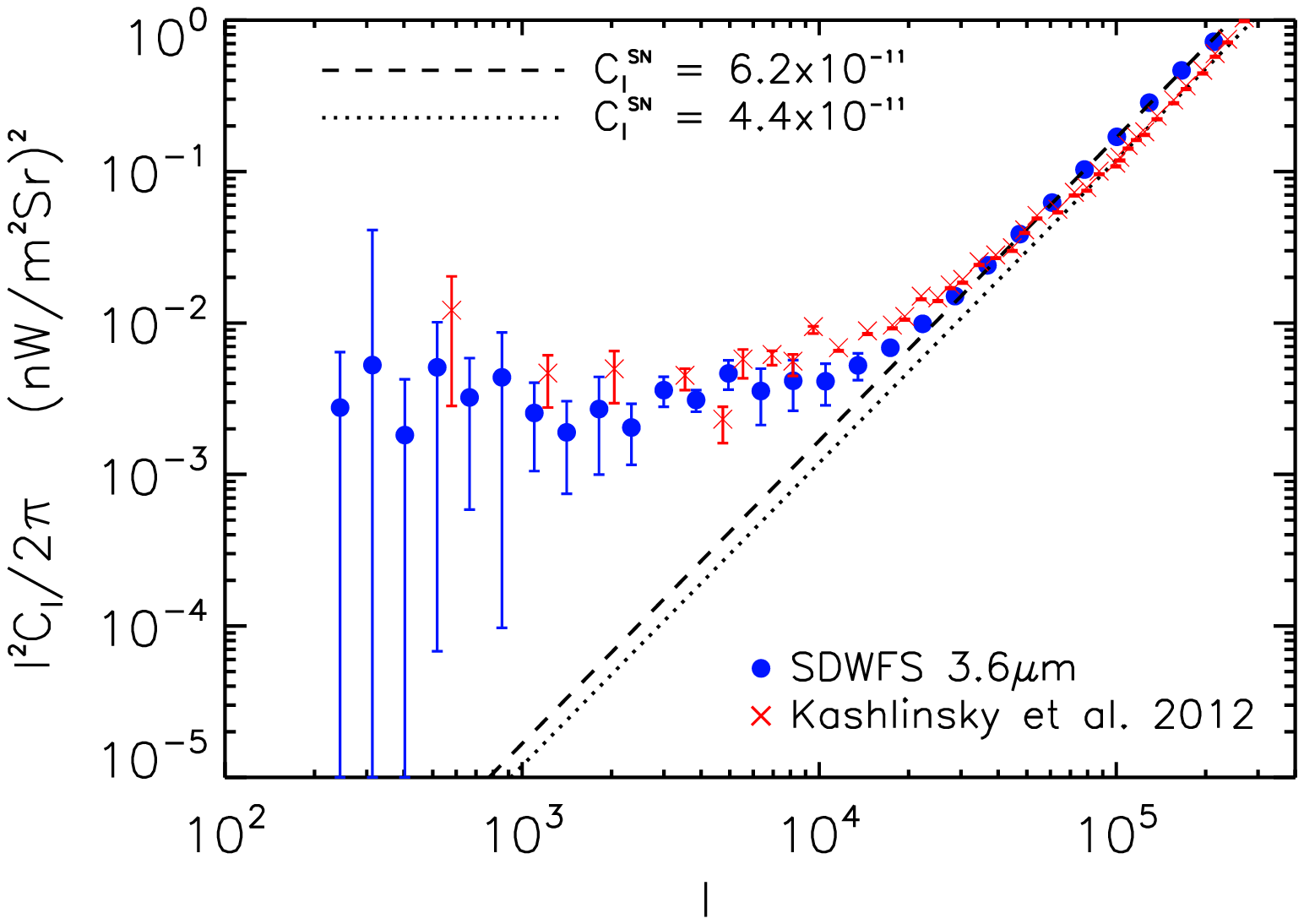},
        \includegraphics[scale=0.7]{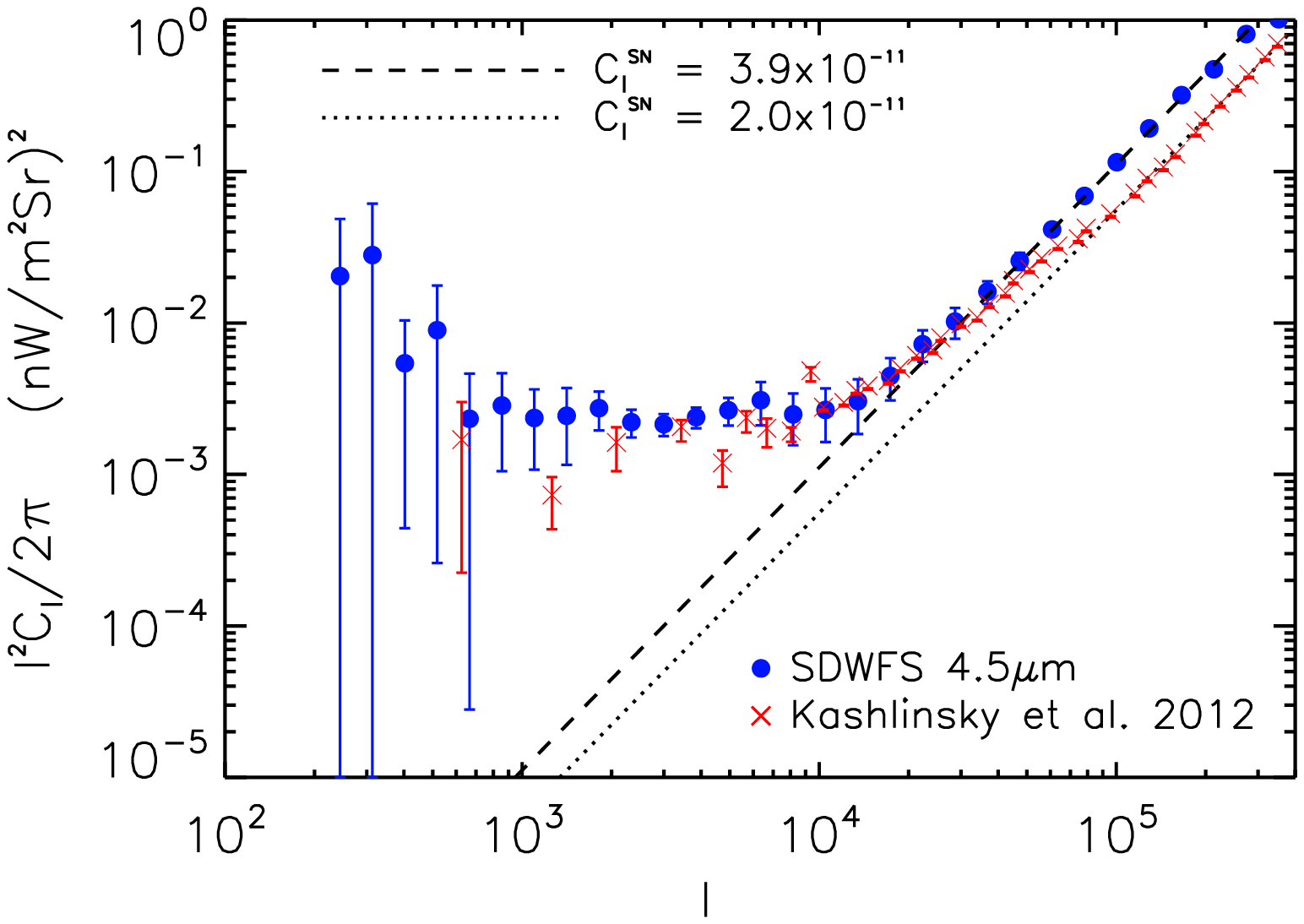}}
    \end{center}
    \caption{{\bf The angular power spectrum of near-IR anisotropies.} The angular power spectrum of near-IR anisotropies measured with SDWFS at 3.6 and 4.5 $\mu$m. The 1 $\sigma$ 
error bars include
all uncertainties we have discussed in the Supplement and the measurements are beam corrected. We also compare our measurements to existing results\cite{Kashlinsky12}
where we find a general agreement on clustering. The large-$\ell$ difference between the two datasets reflect the depth of the point source identification and removal
in the mask.}
   \label{fig:cl}
 \end{figure}

Fig.~S\ref{fig:cl} shows the results from the {\it Spitzer} SDWFS Bo\"otes field (the power spectra values
are listed in Table~\ref{tab:cl_vals}). Results from a recent
analysis\cite{Kashlinsky12} are also given for reference.
These spectra are the final spectra after all corrections have been applied. We note the strong agreement between our measurements and the previously published ones.
The difference at small angular scales, high $\ell$, is due to differences in the depth of the mask. It is captured by a difference of the shot-noise
levels in the point source detection level between SDWFS and deeper SEDS data\cite{Kashlinsky12}.

\begin{figure}
    \begin{center}
        {\includegraphics[scale=0.7]{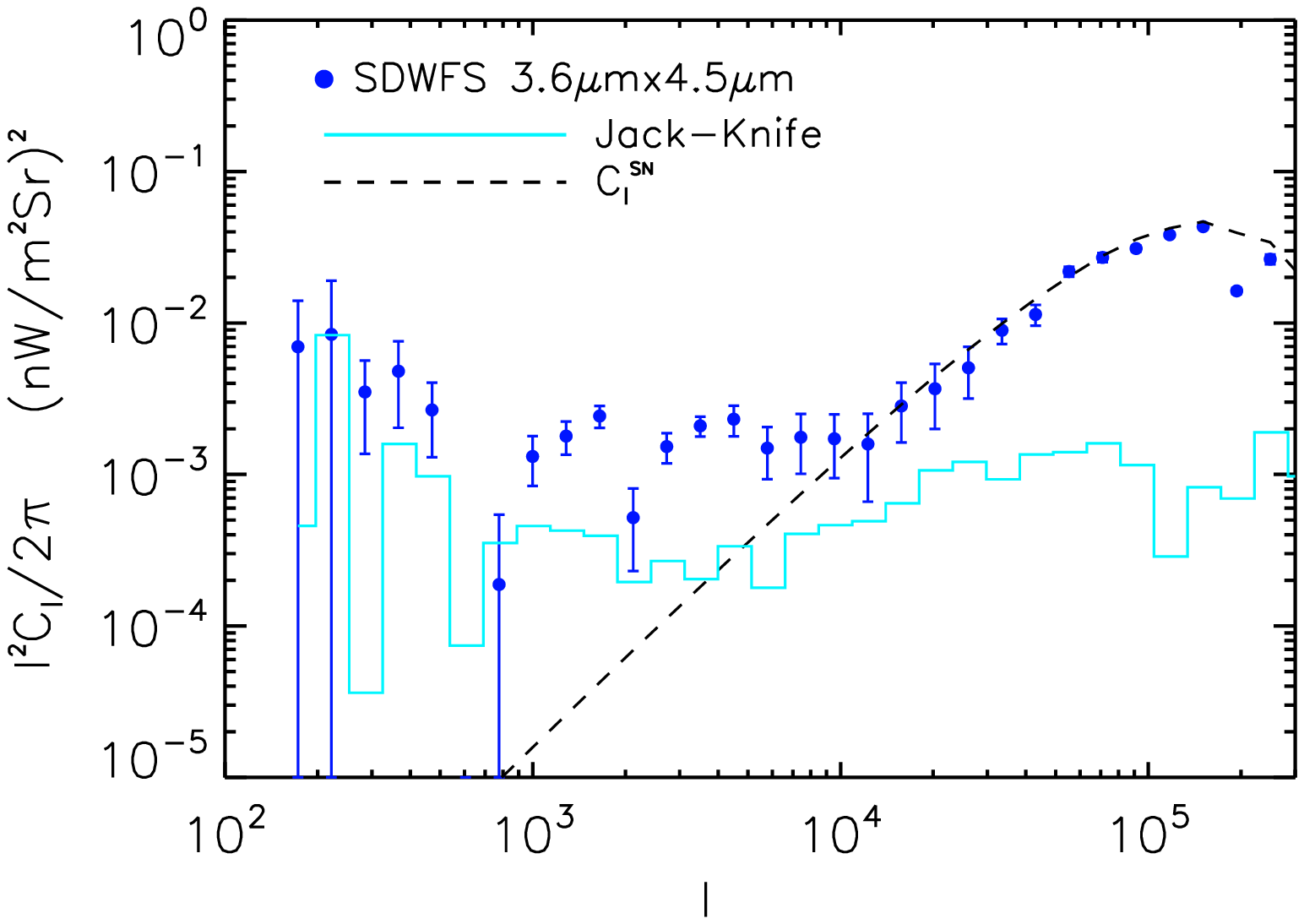},
        \includegraphics[scale=0.7]{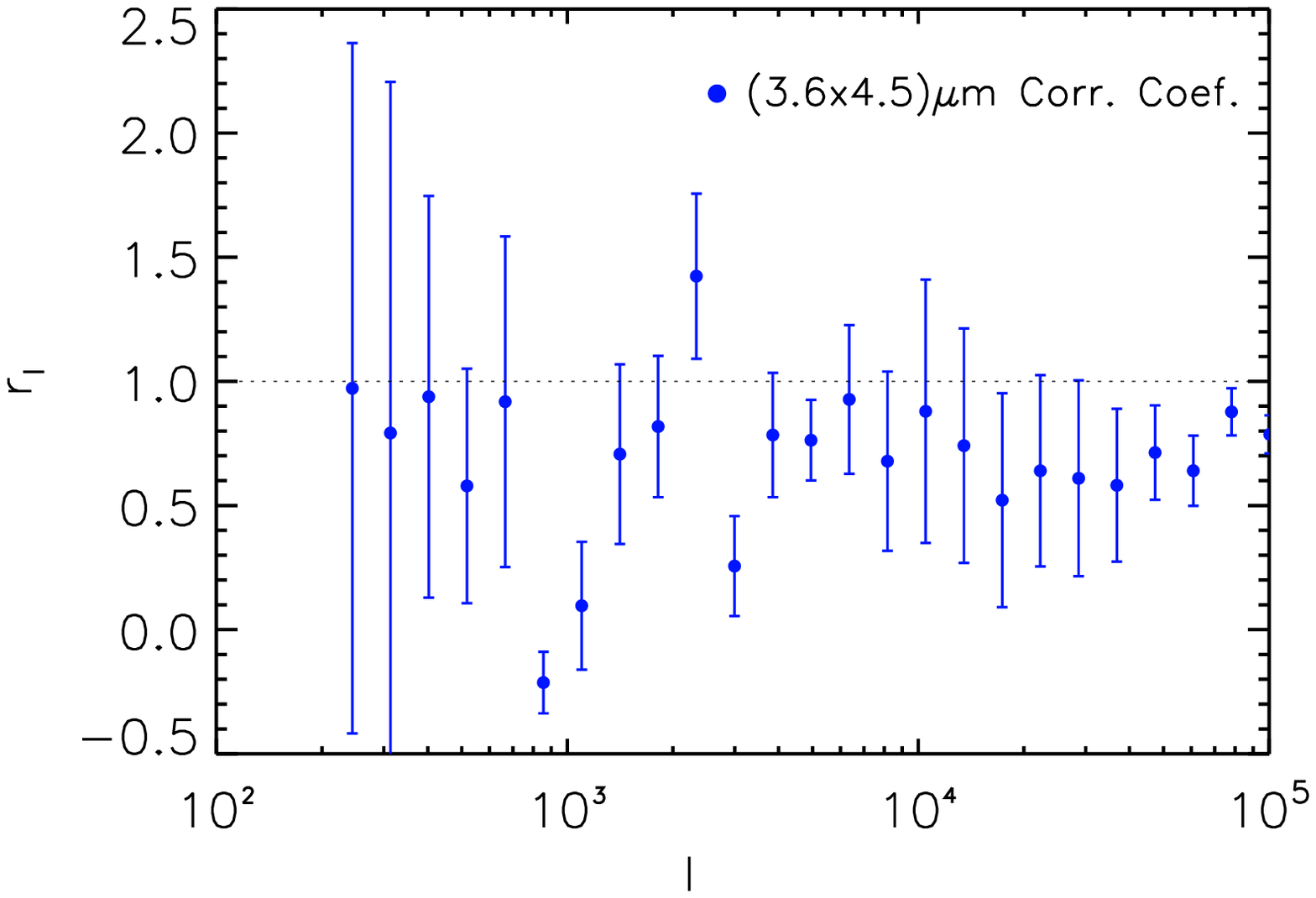}}
    \end{center}
    \caption{{\bf The angular cross-power spectrum of near-IR anisotropies.} The angular cross-power spectrum of near-IR anisotropies measured with SDWFS at 3.6 and 4.5 $\mu$m. 
The 1 $\sigma$ error bars include
all uncertainties we have discussed in the Supplement and the measurements are beam corrected. The upper panel shows the cross power spectrum ($C_l^{3.6-4.5}$), while the lower
panel shows the correlation coefficient calculated as $r=C_l^{3.6-4.5}/\sqrt{C_l^{3.6}C_l^{4.5}}$, where $C_l^{3.6}$ and $C_l^{4.5}$ are the auto power spectra
at 3.6 and 4.5 $\mu$m, respectively (Fig.~S\ref{fig:cl})}.
   \label{fig:cross}
 \end{figure}

Fig.~S\ref{fig:cross} shows the  angular cross-power spectrum of near-IR anisotropies measured with SDWFS at 3.6 and 4.5 $\mu$m. 
The left panel shows the cross power spectrum ($C_l^{3.6-4.5}$) between the two channels, while the right
panel shows the correlation coefficient calculated as 
\begin{equation}
r=C_l^{3.6-4.5}/\sqrt{C_l^{3.6}C_l^{4.5}} \, ,
\label{eq:coeff}
\end{equation} 
where $C_l^{3.6}$ and $C_l^{4.5}$ are the auto power spectra
at 3.6 and 4.5 $\mu$m, respectively (Fig.~S\ref{fig:cl}). The
correlation coefficient is consistent with unity. The  errors are the 1$\sigma$ overall 
uncertainty in the correlation coefficients found by propagating errors on the cross power spectrum and auto power spectra through equation~\ref{eq:coeff}.

\begin{figure}[th!]
    \begin{center}
      \includegraphics[scale=0.6]{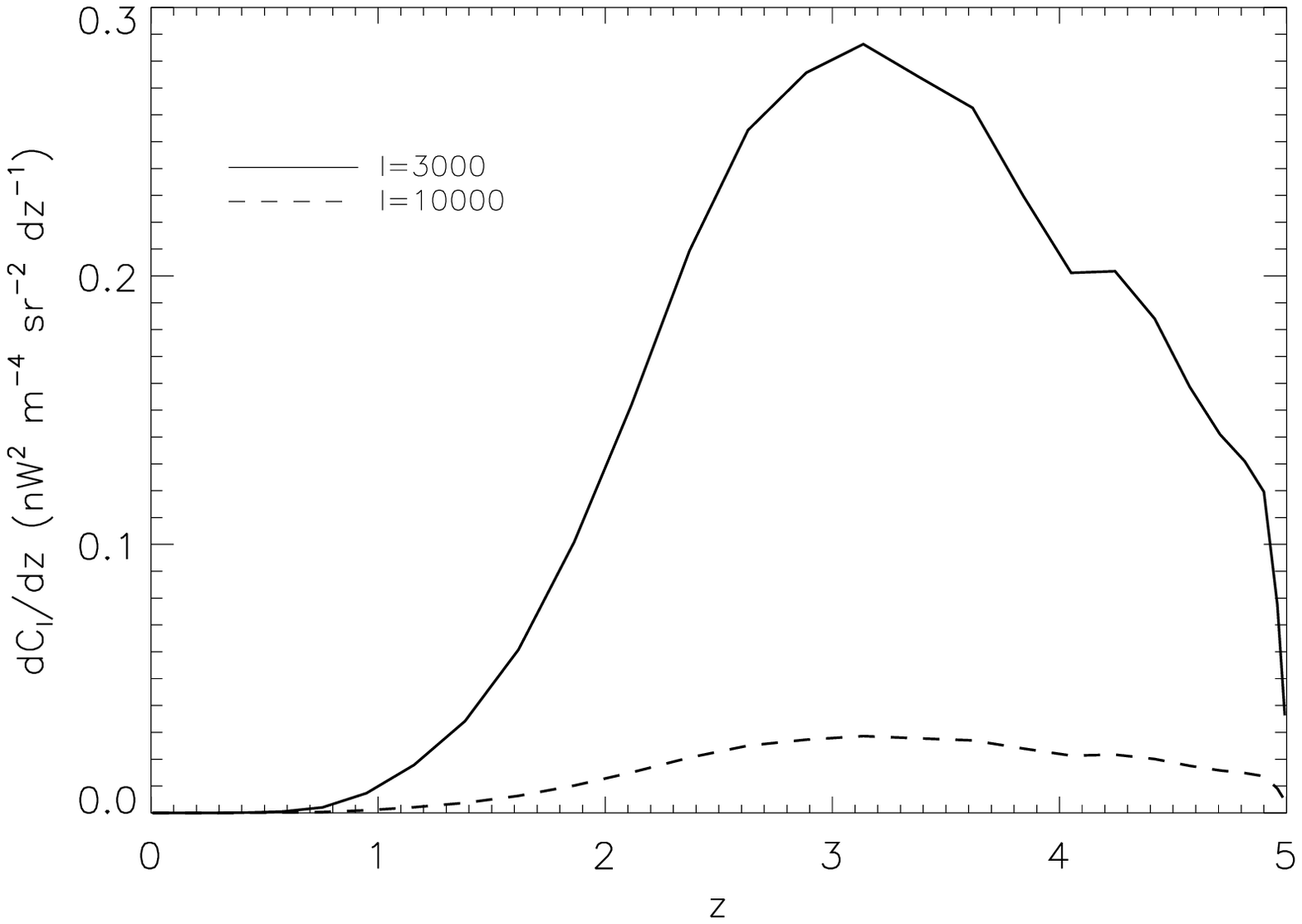}
   \end{center}
   \caption{{\bf The redshift dependence of the IHL anisotropy power spectrum.} $dC_l/dz$ as a function of redshift for $\ell=3\times10^3$ and 10$^4$. The majority of near-IR anisotropies originate from $1 < z < 4$.}
    \label{fig:dcdz}
\end{figure}


\section{Theoretical Interpretation of Near-IR Anisotropies as Spatial Fluctuations of Integrated Intrahalo Light}

Our intrahalo light (IHL) model presented in the {\it Letter} is described in this Section. In Fig.~1 of the main {\it Letter},
we also show results from two descriptions related to the near-IR background anisotropies. One involves the faint galaxies that fall below the
magnitude cut-off of the masks that are applied to measure fluctuations. 
The contribution from faint galaxies, primarily dwarf has been studied
in detail with latest information on the faint-end of the galaxy
luminosity functions\cite{Helgason12} and we use their results in Fig.~1. 
A second model involves the $z > 6$ galaxy contribution. The shaded region for $z > 6$ galaxies in Fig.~1 combines the analytical models\cite{Cooray12} with results from numerical simulations\cite{Fernandez12}. The predictions are normalized to the measured luminosity functions of galaxies at $z > 6$ 
and uses a reionization history that is consistent with the WMAP 7-year optical depth to electron scattering\cite{Cooray12}.

For the interpretation presented in the {\it Letter},  we model the IHL intensity angular power spectrum using the halo model.
The IHL model differs from galaxy clustering models in that we assign a profile to the diffuse stars. The standard galaxy clustering
models assume a central galaxy at the halo center and satellites that are distributed randomly in the halo with a profile that is tracing the dark
matter distribution. We now add a diffuse extended component  in addition to the central and satellite 
galaxies. The halo number density\cite{ST} as a function of redshift and mass $dn(M,z)/dM$ is
\begin{equation}\label{comov_den}
    \frac{dn}{d\ln M}=\frac{\rho_m}{M}f(\nu)\frac{d\nu}{d\ln M}\,,
\end{equation}
 with
\begin{eqnarray}
\nu f(\nu) = A\sqrt{{2 \over \pi} a\nu^2 } [1+(a\nu^2)^{-p}] \exp[-a\nu^2/2]\,.
\end{eqnarray}
Given a galaxy luminosity-halo mass relation, a certain fraction $f_{ICL}(M)$ of this luminosity will be in the form of IHL. The IHL luminosity-mass relation is then:
\begin{eqnarray}
L_{IHL,\lambda}(M,z)=f_{IHL}(M)L(M,z=0)(1+z)^{\alpha}f_\lambda(\lambda/(1+z)) \, ,
\end{eqnarray}
where $\alpha$ is the power-law index that accounts for a possible redshift evolution and $f_\lambda(\lambda/(1+z))$ is the spectral energy distribution
of the IHL (see also discussion below). We model the  fraction of total luminosity in form of IHL as a power-law in halo mass,
\begin{eqnarray}
f_{IHL}(M)=A_f\left(\frac{M}{10^{12}M_{\odot}}\right)^{\beta} \, .
\end{eqnarray}

The total luminosity as a function of halo-mass $L(M,z=0)$ at $z=0$  is taken  to be the best-fit relation from\cite{Lin:2004ak} at $2.2$ $\mu$m
\begin{eqnarray}
L(M,z=0)=5.64\cdot10^{12}h_{70}^{-2}\left(\frac{M}{2.7\cdot10^{14}h^{-1}_{70}M_{\odot}}\right)^{0.72}L_{\odot} \, .
\end{eqnarray}
This is measured for galaxy groups and clusters. At smaller mass scales one no longer has the issue of multiple galaxies in a halo and the
total luminosity is simply that of the central galaxy\cite{Vale}.
We extend it to lower masses using the same power-law slope since analyses of the total luminosity-halo mass relation using galaxy-galaxy lensing find the slope continues down to mass scales below 10$^{11}$ M$_{\odot}$\cite{CoorayMilos,Yang}.
Given that this total luminosity-halo mass relation is measured at 2.2 $\mu$m, we scale this to other wavelengths with the SED $f_\lambda$, 
but with the normalization that $f_\lambda=1$ at 2.2 $\mu$m at $z=0$.
We consider the SED of IHL to be consistent with that of old elliptical galaxies comprised of old, red stars\cite{Krick:2007fp}.
In the main {\it Letter}, in Fig.~3, we also consider alternatives for the SED using a variety of galaxy SED templates.

Under these assumptions the angular power  spectrum  of the IHL flux fluctuations can be written as the sum of a 1-halo term, that originates from small-scale fluctuations within
individual halos
\begin{eqnarray}\label{1halo}
C_{\ell}^{1h}=\frac{1}{(4\pi)^2}\int dV \frac{1}{(1+z)^2\chi^4(z)}\int_{M_{min}}^{M_{max}}dM\frac{dn(M,z)}{dM}u_{\rm IHL}^2(k|M)L_{IHL,\lambda}^2(M,z) \, ,
\end{eqnarray}
  and a 2-halo term, related to the large scales dark matter fluctuations, and hence to the linear dark matter power spectrum $P_{M}(k,z)$ as
\begin{eqnarray}\label{2halo}
C_{\ell}^{2h}& =&\frac{1}{(4\pi)^2}\int dV\frac{1}{(1+z)^2\chi^4(z)}\left[\int_{M_{min}}^{M_{max}} dM\frac{dn(M,z)}{dM}u_{\rm IHL}(k|M)b_h(M,z)L_{IHL,\lambda}(M,z)\right]^2\nonumber \\
&& \times P_{M}(k=\ell/\chi(z),z) \, ,
\end{eqnarray}
where $u_{\rm IHL}(k,z|M)$ is the Fourier transform of the IHL profile in a dark matter halo of mass $M$ at redshift $z$, $b_h(M)$ is the linear bias,
 $\chi(z)$ is the comoving radial distance, and $dV$ is the comoving volume element $dV=\chi(z)^2d\chi/dz$. 
The redshift integration is performed up to a maximum redshift $z_{\rm max}=5$. We found that integrating to a higher redshift did not change our results.
The values of $M_{min}$ and $M_{max}$ in Eqs.~\ref{1halo} and \ref{2halo} determine the relative amplitude of the 1-halo and 2-halo terms and we let them vary freely.
The power spectrum also contains a shot-noise contribution from unresolved fluctuations, so that the total power spectrum is
\begin{eqnarray}
C_{\ell}=C_{\ell}^{1h}+C_{\ell}^{2h}+C^{\rm SN}_{\ell} \,.
\end{eqnarray}
We also take $C^{\rm SN}_{\ell}$ to be a free parameter that is varied during our model fitting.

Since the IHL profile for small mass halos has yet to be determined precisely, we consider two model descriptions under the assumption that
IHL (i) traces the Navarro-Frenk-White (NFW) profile of dark matter halos\cite{NFW} and (ii) falls as $r^{-2}$ with an exponential cut-off\cite{Masaki}
such that $\rho_{\rm IHL}  \propto 1/r^2 \exp(-r/2R_{\rm vir})$. There is limited information on the light profile from the
stacking analysis of SDSS galaxies\cite{Tal}. However, we are unable to use those measurements for the
current study as we do not have information on how the profile changes with the halo mass.
Moreover given the limited information both in terms of the angular scale of fluctuations and the large uncertainties we find that we are not able to
statistically distinguish one IHL profile over another. 

In Fig.~1 of the main {\it Letter}, the best-fit anisotropy power spectrum makes use of the description involving the NFW profile with
\begin{eqnarray}
\rho(r)=\frac{\rho_s}{(cr/r_{\rm vir})(1+cr/r_{\rm vir})^2}\,,
\end{eqnarray}
where $r_{vir}$ is the virial radius and c the halo concentration parameter. We define the concentration using the result from numerical
simulations\cite{Bullock1999he} that find
\begin{eqnarray}
c(M,z)&=&\frac{9}{1+z}\left(\frac{M}{M_*}\right)^{-0.13}\,,
\end{eqnarray}
where $M_*$ is the mass scale at which the critical density contrast $\delta_c$ required for spherical collapse is equal to the square root of the variance in the initial density field $\sigma(M_*)=\delta_c$.  While we make use of this particular fitting function\cite{Bullock1999he}, 
an alternative fitting function\cite{Duffy08} led to the same results. Thus, our best-fit parameter values are independent of the assumption
on the halo mass-concentration relation.

To analyze the data we conducted a Monte Carlo-Markov Chain (MCMC) analysis using a modified version of the publicly available package cosmoMC\cite{Lewis:2002ah} with convergence diagnostic based on the Gelman-Rubin statistic\cite{gelmanrubin}. We fit a total of six 
free parameters: the minimum and maximum masses ($M_{\rm min}$, $M_{\rm max}$), 
the power-law index of the redshift dependence in the luminosity-mass relation $\alpha$, the amplitude and the power-law index with halo mass 
of the IHL fraction $A_f$ and $\beta$, respectively, and the shot noise contribution 
$C^{\rm SN}_\ell$. We fix the cosmological parameters to the best-fit for the $\Lambda CDM$ concordance model from WMAP 7-year analysis\cite{Komatsu:2010fb}. 

The best-fit parameters for the fit to the data at 3.6 $\mu$m from SDWFS are shown in Table~\ref{parameters}.  The minimum and maximum halo masses
of the halo mass range contributing to measured near-IR anisotropies is 10$^{9.03 \pm 0.05}$ M$_{\odot}$ and 10$^{11.91 \pm 0.05}$ M$_{\odot}$, 
respectively. Our model-fitting suggests a small dependence of the IHL fraction on halo mass with a power-law slope of $\beta=0.09 \pm 0.01$.
We found values consistent with the 1$\sigma$ uncertainties when using the alternate IHL profile described above and fitting the model to the 4.5 $\mu$m
power spectrum measurements. Our results for $f_{\rm IHL}(M)$ are
summarized in Fig.~2 of the main {\it Letter}. There we also compare our determination to an analytical prediction of the IHL fraction relative to the total luminosity, in the literature\cite{Purcell08}. These analytical models exist only down to a 
halo mass scale of about $5 \times 10^{10}$ M$_{\odot}$ and $f_{\rm IHL} > 5\times10^{-4}$.
In addition to the power-law behavior it was found also found in previous analytical work\cite{Purcell07} that the IHL fraction is constant
at a value of about 0.005 when $M < 5 \times 10^{11}$ M$_{\odot}$ in some scenarios to generate IHL.
 This flattening behavior could be related to our observation that the
IHL fraction is not strongly halo mass dependent over the mass ranges that we are probing with near-IR background anisotropy power spectrum. More detailed studies are necessary to properly understand how our results can be used to understand the merger rate
and generation of IHL in low mass halos at redshifts of 1 to 4.

In Fig.~S\ref{fig:dcdz} we summarize the redshift dependence of the IHL
power spectrum by calculating $dC_l/dz$ as a function of redshift. The contributions peak at a redshift of 3, but has a broad distribution ranging from $1 < z < 4$. 
The measured shot-noise level of $120 \pm 10$ nJy nW m$^{-2}$ sr$^{-1}$ is about a factor of 2 higher than the shot-noise in the
deeper SEDS data\cite{Kashlinsky12} with a value of $\sim 57$ nJy nW m$^{-2}$ sr$^{-1}$.

In Fig.~3 of the main {\it Letter}, we summarize the  SED of IHL. Here, we make use of a variety of stellar templates from B to K-type stars for this purpose.
The prediction converges at the longer wavelengths due to existing measurements but we find large deviations at 1 $\mu$m and shorter wavelengths.
A measurement of the background anisotropy at optical wavelengths is clearly desirable and could be used to both identify the
stars that are primarily contributing to IHL at $z \sim$1 to 4.

\end{document}